\documentclass[trackchanges, twocolumn]{aastex701}
\newcommand{\fion}[2]{\mbox{[#1\,\uppercase\expandafter{\romannumeral #2\relax}]}}
\newcommand{\sfion}[2]{\mbox{#1\,\uppercase\expandafter{\romannumeral #2\relax}]}}

\newcommand{\appropto}
{\mathrel{\vcenter{\offinterlineskip\halign{\hfil$##$\cr\propto\cr\noalign{\kern2pt}\sim\cr\noalign{\kern-2pt}}}}}
\newcommand{\bagpipes}{{\tt{Bagpipes}}}
\newcommand{\prospector}{{\tt{Prospector}}}

\newcommand{\sextractor}{{\tt{SExtractor}}}

\newcommand{\eazypy}{{\tt{EaZy-py}}}
\newcommand{\galfind}{{\tt{GALFIND}}}
\newcommand{\jwst}{\emph{JWST}}
\newcommand{\hst}{\emph{HST}}
\newcommand{\webbpsf}{{\tt{WebbPSF}}}
\newcommand{\MUV}{$M_{\mathrm{UV}}$}

\newcommand{\bpass}{{\tt{BPASS}}}

\newcommand{\cloudy}{{\tt{CLOUDY}}}

\newcommand{\epochs}{{\tt EPOCHS}}
\newcommand{\beagle}{{\tt BEAGLE}}

\newcommand{\BDFinder}{{\tt BD-Finder}}

\newcommand{\lya}{Ly$\alpha$}
\newcommand{\ha}{H$\alpha$}
\newcommand{\hbeta}{H$\beta$}

\newcommand{\xiion}{$\xi_{\rm ion}$}
\newcommand{\xiionzero}{$\xi_{\rm ion, 0}$}
\newcommand{\Ndotion}{$\dot{N}_{\rm ion}$}
\newcommand{\ndotion}{$\dot{n}_{\rm ion}$}

\newcommand{\Ndotionzero}{$\dot{N}_{\rm ion, 0}$}
\newcommand{\fesc}{$f_{\rm esc}^{\rm LyC}$}
\newcommand{\cHII}{$C_{\rm HII, rec}$}
\newcommand{\dQdtzero}{$\mathrm{d}Q_{\rm HII}/\mathrm{d}t=0$}
\newcommand{\dndotdz}{$\mathrm{d}\log_{10}\dot{n}_{\rm ion}/\mathrm{d}z$}

\usepackage{subcaption}
\usepackage{amsmath}
\usepackage{multirow}
\usepackage{placeins} 

\shorttitle{IGM clumping constraints}
\shortauthors{Duncan Austin et al.}

\begin{document}

\title{Resolving the ionizing photon budget crisis with JWST/NIRCam \ion{H}{2} clumping constraints at \boldmath\texorpdfstring{$z\simeq6$}{z=6}}


\author[0000-0003-0519-9445]{Duncan Austin}
\affiliation{Jodrell Bank Centre for Astrophysics, Alan Turing Building, University of Manchester, Oxford Road, Manchester M13 9PL, UK}
\email[show]{duncan.austin@manchester.ac.uk}

\author[0000-0002-4130-636X]{Thomas Harvey}
\affiliation{Jodrell Bank Centre for Astrophysics, Alan Turing Building, University of Manchester, Oxford Road, Manchester M13 9PL, UK}
\email[]{thomas.harvey-3@manchester.ac.uk}

\author[0000-0003-1949-7638]{Christopher J. Conselice}
\affiliation{Jodrell Bank Centre for Astrophysics, Alan Turing Building, University of Manchester, Oxford Road, Manchester M13 9PL, UK}
\email[]{conselice@manchester.ac.uk}

\author[0000-0003-4875-6272]{Nathan J. Adams}
\affiliation{Jodrell Bank Centre for Astrophysics, Alan Turing Building, University of Manchester, Oxford Road, Manchester M13 9PL, UK}
\email[]{nathan@theadamshouse.co.uk}

\author[0000-0001-7633-3985]{Vadim Rusakov}
\affiliation{Jodrell Bank Centre for Astrophysics, Alan Turing Building, University of Manchester, Oxford Road, Manchester M13 9PL, UK}
\email[]{vadim.rusakov@manchester.ac.uk}

\author[0000-0002-3119-9003]{Qiong Li}
\affiliation{Jodrell Bank Centre for Astrophysics, Alan Turing Building, University of Manchester, Oxford Road, Manchester M13 9PL, UK}
\email[]{qiong.li@manchester.ac.uk}

\author[0009-0008-8642-5275]{Lewi Westcott}
\affiliation{Jodrell Bank Centre for Astrophysics, Alan Turing Building, University of Manchester, Oxford Road, Manchester M13 9PL, UK}
\email[]{lewi.westcott@manchester.ac.uk}

\author[0009-0006-4799-4497]{Caio Goolsby}
\affiliation{Jodrell Bank Centre for Astrophysics, Alan Turing Building, University of Manchester, Oxford Road, Manchester M13 9PL, UK}
\email[]{caio.goolsby@postgrad.manchester.ac.uk}

\author[]{Kai Madgwick}
\affiliation{Jodrell Bank Centre for Astrophysics, Alan Turing Building, University of Manchester, Oxford Road, Manchester M13 9PL, UK}
\email[]{kyle.madgwick@postgrad.manchester.ac.uk}

\author[]{James Arcidiacono}
\affiliation{Jodrell Bank Centre for Astrophysics, Alan Turing Building, University of Manchester, Oxford Road, Manchester M13 9PL, UK}
\email[]{james.arcidiacono@manchester.ac.uk}

\author[0000-0003-4223-7324]{Massimo Ricotti}
\email{ricotti@astro.umd.edu}
\affiliation{Department of Astronomy, University of Maryland, College Park, 20742, USA}

\author[0009-0001-3422-3048]{Sophie L. Newman}
\affiliation{Institute of Cosmology and Gravitation, University of Portsmouth , Burnaby Road, Portsmouth PO1 3FX, UK}
\email[]{sophie.newman@port.ac.uk}

\author[0000-0002-7020-3079]{Louise T. C. Seeyave}
\affiliation{Astronomy Centre, University of Sussex, Falmer, Brighton BN1 9QH, UK}
\email[]{louise.seeyave@gmail.com}

\author[0000-0002-9081-2111]{James Trussler}
\affiliation{Center for Astrophysics, Harvard \& Smithsonian, 60 Garden St., Cambridge MA 02138, USA}
\email[]{james.trussler@cfa.harvard.ac.uk}

\author[0000-0003-1625-8009]{Brenda Frye}
\email[]{brendafrye@gmail.com}
\affiliation{Department of Astronomy/Steward Observatory, University of Arizona, 933 N Cherry Ave, Tucson, AZ, 85721-0009, USA}


\author[0000-0001-9440-8872]{Norman A. Grogin}
\email[]{nagrogin@stsci.edu}
\affiliation{Space Telescope Science Institute, 3700 San Martin Drive, Baltimore, MD 21218, USA}



\author[0000-0003-1268-5230]{Rolf A. Jansen}
\email{rolfjansen.work@gmail.com}
\affiliation{School of Earth and Space Exploration, Arizona State University, Tempe, AZ 85287-1404, USA}

\author[0000-0002-6610-2048]{Anton M. Koekemoer}
\affiliation{Space Telescope Science Institute, 3700 San Martin Drive, Baltimore, MD 21218, USA}
\email[]{koekemoer@stsci.edu}




\author[0000-0003-3382-5941]{Nor Pirzkal}
\email[]{npirzkal@stsci.edu}
\affiliation{Space Telescope Science Institute, 3700 San Martin Drive, Baltimore, MD 21218, USA}


\author[0000-0001-7016-5220]{Michael Rutkowski}
\email[]{michael.rutkowski@mnsu.edu}
\affiliation{Minnesota Institute for Astrophysics, Minnesota State University, Mankato, MN 56001-6062, USA}




\author[0000-0001-8156-6281]{Rogier A. Windhorst}\email{Rogier.Windhorst@gmail.com}
\affiliation{School of Earth and Space Exploration, Arizona State University, Tempe, AZ 85287-1404, USA}




\begin{abstract}

We present a comprehensive study of the ionizing properties of 1721 galaxies at $5.6<z<6.5$ using deep JWST/NIRCam photometric imaging from the NEP, JADES, and PRIMER surveys spanning an unmasked area $\sim550$~arcmin$^2$ across UV magnitudes $-22\lesssim M_{\rm UV}\lesssim-17.5$. Our 90\% stellar mass complete sample suggests little relation of UV slope with magnitude, $\beta_{\rm UV}=(-0.040\pm0.022)M_{\rm UV}-2.88^{+0.43}_{-0.44}$, implying \fesc$\simeq5\%$ based on calibrations from the Low-redshift Lyman Continuum Survey (LzLCS). We measure a constant ionizing photon production efficiency with UV magnitude, $\log_{10}(\xi_{\rm ion, 0}/\rm Hz\,erg^{-1}) = -0.006^{+0.019}_{-0.017}~M_{\rm UV} + 25.05^{+0.39}_{-0.34}$, consistent with HST canonical values. The total production rate of photons escaping into the IGM is computed as $\log_{10}(\dot{n}_{\rm ion}/\rm s^{-1}Mpc^{-3})=50.31^{+0.07}_{-0.06}$, for \MUV\,$<-17$ galaxies from our star forming and smouldering UV luminosity functions (UVLFs), which differ in the faint-end slope ($\alpha_{\rm SFG}=-2.2\pm0.2$; $\alpha_{\rm sm}=-1.7\pm0.2$). Extrapolating to the latest UVLF turnover limits from the massive lensing galaxy cluster Abell S1063 ($M_{\rm UV, lim}=-13.5$) implies that a recombination-weighted \ion{H}{2} clumping factor \cHII\,$=6.2^{+4.1}_{-2.1}$ is required to produce fully stably reionized at $z\simeq6$. A clumping factor of this magnitude resolves the ionizing photon budget crisis. Our methodology paves the way for indirect clumping measurements from galaxies which will provide insight into earlier stages of the EoR when the \lya-forest becomes saturated and more direct quasar measurements become impossible.

\end{abstract}



\section{Introduction}
\label{sec:introduction}

The Epoch of Reionization (EoR) is the period between the formation of the first stars and galaxies and the complete ionization of the previously neutral intergalactic medium (IGM) by Lyman continuum (LyC) photons. Its midpoint is constrained to be $z=7.7\pm 0.7$ \citep{Planck2020}, and observations of the Gunn-Peterson (GP) trough \citep[][]{Gunn-Peterson1965} and damping wings in quasar spectra define the end of the hydrogen EoR to be $z\sim5-6$ \citep[e.g.][]{Keating2020, Bosman2022, Keating2024, Zhu2024}.


While the contribution of active galactic nuclei (AGN) to the total production rate of photons ionizing the IGM remains unclear \citep{Maiolino2024, Lambrides2024, Jiang2025}, simulations have shown that their contribution is subdominant \citep[$\lesssim10-30\%$;][]{Yung2021, Asthana2024, Dayal2025} and that massive, short lived O/B main-sequence stars dominate the total production of ionizing photons from galaxies \citep[e.g.][]{Rosdahl2018, Trebitsch2021}. In the early Universe, it has been proposed that globular clusters are the dominant source \citep{Ricotti2002, He2020}. To determine the total rate of photon injection into the IGM from star-forming galaxies, \ndotion, three main ingredients are required: 1) the UV luminosity function (UVLF; $\Phi_{\rm UV}$); 2) the rate (\Ndotion) and efficiency (\xiion) of ionizing photon production; and 3) measurement of the LyC escape fraction, \fesc, defined as the fraction of LyC photons escaping the interstellar medium (ISM) of their host galaxies.

The UVLF in the EoR at $z>6.5$ was measured using deep photometric imaging with the Hubble Space Telescope \citep[HST;][]{Oesch2014, Bouwens2015, Finkelstein2015, McLeod2015, McLeod2016, Oesch2018} and more recently with JWST/NIRCam \citep[e.g.][]{Donnan2023, Harikane2023, Adams2024, Donnan2024, McLeod2024} down to $M_{\rm UV}\simeq -17/18$ mag. Promising future avenues exploit the magnification from strong-lensing clusters in the Ultradeep NIRSpec and NIRCam Observations before the Epoch of Reionization survey \citep[UNCOVER; PIs: I. Labbe \& R. Bezanson; PID: 2561;][]{Bezanson2024} and Gravitational Lensing \& NIRCam IMaging to Probe early galaxy formation and Sources of rEionization \citep[GLIMPSE; PIs: H. Atek \& J. Chisholm; PID: 3293;][]{Atek2025} programs. These surveys probe fainter magnitudes down to $M_{\rm UV}\simeq-12$, albeit over increasingly small areas, and even though photometric candidates have been found down to $M_{\rm UV}\sim-13.5$, for example in the Sunrise Arc \citep{Vanzella2023}, no strong evidence for a turnover has yet been found.


The ionizing photon production efficiency is an intrinsic galaxy property which is commonly calculated from the ratio of either the \fion{O}{3}$~\lambda5007$ \citep{Nakajima2016, Tang2019, Izotov2021a} or Balmer recombination lines with the UV luminosity, tracing the burstiness of star formation. Canonical values are taken to be $\log_{10}(\xi_{\rm ion}/\rm Hz~erg^{-1})=25.2$ supported by measurements from the Hubble Ultra Deep Field \citep[HUDF;][]{Robertson2013}, Spitzer/IRAC at $3.6~\mu \rm m$ and $4.4~\mu \rm m$ for $z=4-5$ galaxies \citep{Bouwens2016b}, and the MOSFIRE Deep Evolution Field \citep[MOSDEF;][]{Kriek2015, Reddy2018}, although attenuation by dust may reduce observed values by $\sim0.2-0.3~\rm dex$ \citep{Shivaei2018}. Several studies have found an increasing trend of \xiion\ with redshift \citep[e.g.][]{Finkelstein2019, Stefanon2022, Begley2025a}, where harder ionizing spectra and burstier star formation histories (SFHs) are expected. In addition, larger $\xi_{\rm ion}$ values have been observed in \lya\ emitters \citep[LAEs;][]{Matthee2017, Harikane2018, Maseda2020, Chen2024a, Saxena2024}. Much current discussion in the literature surrounds the burstiness of UV luminous sources at $z\gtrsim6$, with several studies suggesting a positive trend in $\xi_{\rm ion}-M_{\rm UV}$ such that fainter galaxies have larger ionizing photon production efficiencies \citep[e.g.][]{Prieto-Lyon2023, Simmonds2024a, Llerena2025a}. More recent studies, however, suggest that the enhanced selectability of faint bursty galaxies biases this trend, and that the underlying distribution is either flatter or negatively sloped such that brighter galaxies have larger \xiion\ values \citep[e.g.][]{Endsley2024, Begley2025a, Pahl2025a, Simmonds2024b}.

The fraction and escape mechanisms of ionizing photons from the ISM and into the IGM is widely debated. Theories include escape from density bounded nebulae where small ISM neutral gas reservoirs become saturated by the large amount of ionizing photons \citep{Zackrisson2013, Nakajima2014, Katz2020} or escape via holes punched through the ISM either by outflows of stellar or AGN origin \citep{Gazagnes2020, Katz2023}. Either way, these \fesc\ values are challenging to measure in the EoR at $z\gtrsim6$ due to the complete absorption of LyC photons by the ionized IGM along the line-of-sight. Promising indirect tracers of \fesc\ exist, including from the Low-redshift Lyman Continuum Survey \citep[LzLCS;][]{Flury2022a, Flury2022b}, where $\beta_{\rm UV}$ measurements are seen to correlate with \fesc\ \citep{Chisholm2022}, albeit with a large scatter. In addition to $\beta_{\rm UV}$, trends of escape fraction have been observed with the \fion{Mg}{2}-$\lambda\lambda2796,2804$ doublet \citep{Henry2018, Chisholm2020, Xu2022, Katz2022}, $O_{32}$ ratios \citep[\fion{O}{3}~$\lambda5007$/\fion{O}{2}~$\lambda\lambda3726,3728$, e.g.][]{Jaskot2013, Nakajima2014, Izotov2018a, Izotov2018b}, \lya\ velocity offset \citep{Verhamme2015, Verhamme2017, Izotov2020} or equivalent width \citep[EW;][]{Steidel2018, Pahl2021}, \fion{S}{2} deficits \citep{Wang2019, Wang2021}, and SFR surface density \citep[$\Sigma_{\rm SFR}$;][]{Naidu2020}. Tracing \fesc\ in the EoR is clearly a complex procedure, and modern advancements in deep learning methods have allowed for multivariate models to be built based on LzLCS observations at low-redshift \citep{Jaskot2024a, Jaskot2024b} and the SPHINX simulation \citep{Rosdahl2022, Choustikov2024a, Choustikov2024b}.

Even with conservative estimates for the average $f_{\rm esc}^{\rm LyC}\sim10\%$ and $M_{\rm UV, lim}\sim-16$ for SFGs at $z\gtrsim6$, the high $\xi_{\rm ion}$ values measured by JWST imply that the Universe is fully reionized earlier than $z\simeq6$ even without any contribution from AGN, leading to the so-called ``ionizing photon budget problem'' \citep{Munoz2024}. One possible solution to this would be an increasingly clumpy IGM which would reduce the average recombination timescale meaning more ionizing photons are required to stably ionize the Universe at a given ionized hydrogen content, $x_{\rm HII}$ \citep[e.g.][]{Davies2024}. 

In this study, we use deep photometric data from JWST/NIRCam to produce samples of star-forming and smouldering\footnote{Terminology adopted from \citet{Trussler2024}. Note that their definition stems from NIRCam colours probing rest-optical emission lines, and not an estimate of star formation burstiness.} galaxies at $5.6<z<6.5$, split by star-formation burstiness ($\Phi=\allowbreak \rm SFR_{10\rm Myr}/\rm SFR_{\rm 100 Myr}$) as measured by the \bagpipes\ SED fitting code \citep{Carnall2018}. Specifically, our smouldering sample is defined as galaxies with a recent downturn in recent star formation, i.e. $\Phi<1$ or equivalently $\rm SFR_{10\rm Myr}<\rm SFR_{\rm 100 Myr}$. At these redshifts, the optical continuum is traced by F410M (the only medium band common to these blank fields) and the \ha+\fion{N}{2} emission line complex falls within F444W. $\dot{N}_{\rm ion}-M_{\rm UV}$ and $f_{\rm esc}^{\rm LyC}-M_{\rm UV}$ scaling relations, as well as UVLFs, from each subsample are combined to calculate the total ionizing photon production rate, \ndotion.

We assume a flat $\Lambda$CDM cosmology with $H_0=70~\rm km s^{-1}Mpc^{-1}$ , $\Omega_{\Lambda,0}=0.7$ and $\Omega_{\rm m, 0}=0.3$ a helium mass fraction, $Y=0.25$, and present-day cosmic baryon fraction $\Omega_{\rm b, 0}=0.05$. The AB magnitude system \citep{Oke74, Oke83} is used throughout.

\section{Data and sample} 
\label{sec:data_and_sample}

\subsection{{\tt EPOCHS-v2} data and surveys}

In this work, we use deep photometric imaging from both JWST/NIRCam and HST/ACS WFC to produce a robust sample of galaxies at $5.6<z<6.5$. We use blank field JWST/NIRCam data, combined with ancillary HST/ACS WFC imaging, produced using the same \galfind\ pipeline as those in the \epochs\ series \citep{Adams2024, Austin2024, Conselice2025, Harvey2025}, with catalogues and further description of the methods presented in the {\tt EPOCHS-v2} data release (Austin et al. in prep). In short, these catalogues provide forced photometry in $0.32\arcsec$ diameter circular apertures, with aperture corrections derived from \webbpsf\ PSFs \citep{Perrin2012-WebbPSF, Perrin2014-WebbPSF}, with source detection and deblending performed by \sextractor\ \citep{Bertin1996-SExtractor} in a weighted stack of the JWST/NIRCam LW widebands (F277W+F356W+F444W). Photometric uncertainties are determined empirically from the local depth.

Five of the widest and deepest blank field areas with deep JWST/NIRCam photometry are used, covering $550\arcmin^2$ across UV magnitudes $-22\lesssim M_{\rm UV}\lesssim-17.5$. Note that we do not include data from the Cosmic Evolution Early Release Science \citep[CEERS; PID: 13445; PI: S. Finkelstein][]{Bagley2023-CEERS} or Next Generation Deep Extragalactic Exploratory Public \citep[NGDEEP; PID 2079; PIs: Papovich, Pirzkal][]{Bagley2023-NGDEEP} surveys since they do not include the F090W filter crucial for robust photometric redshift estimates at $z\simeq6$.

The faint end is probed by data from the JWST Advanced Deep Extragalactic Survey \citep[JADES;][see also \citet{Bunker2023-NIRSpec} and \citet{Hainline2024a}]{Eisenstein2023} in the Great Observatories Origins Deep Survey (GOODS) North ($78.6\arcmin^2$) and South ($123.5\arcmin^2$) fields, with the UVLF ``knee'' covered by North Ecliptic Pole Time Domain Field \citep[NEP-TDF;][]{Jansen2018, OBrien2024} imaging ($54.5\arcmin^2$) taken from the Prime Extragalactic Areas for Reionization and Lensing Science \citep[PEARLS; PIDs: 1176, 2738; PIs: R. Windhorst \& H. Hammel;][]{Windhorst2023}. The bright end data is dominated by PRIMER (PI: J. Dunlop, PID: 1837) in the Ultra Deep Survey (UDS; $163.7\arcmin^2$) and Cosmological Evolution Survey (COSMOS; $129.3\arcmin^2$). 

Ancillary data from the Hubble Legacy Fields \citep[HLF][]{Illingworth2016, Whitaker2019-HLF} and the Cosmic Assembly Near-IR Deep Extragalactic Legacy Survey \citep[CANDELS][]{Koekemoer2011-CANDELS, Grogin2011-CANDELS} are used to provide optical coverage at $0.4-0.8\mu\rm m$.

\subsection{Sample selection criteria}
\label{sec:EPOCHS_v2_sample_selection}

To robustly select a sample of $5.6<z<6.5$ galaxies, we have modified the EPOCHS selection criteria from \citet{Conselice2025} to account for the fact that, post reionization, the \lya\ break is no longer a step-function at $z\lesssim6.5$; depending on the line-of-sight, some Lyman-series photons may remain unabsorbed by the fully ionized IGM. We run \eazypy\ \citep{Brammer2008-EAZY} with the ``fsps\_larson'' template set \citep{Larson2023} to calculate fiducial photometric redshifts (both with redshift free as well as constrained at $z<4.0$ for low-redshift comparisons), which includes the damped \lya\ (DLA) photo-z correction procedure using the average $N_{\rm HI}$ column densities given in \citet{Asada2025}. A $10\%$ minimum flux error is applied to the photometry during the SED fitting procedure. The sample selection criteria using the photo-z PDFs from \eazypy\ are as follows.

\begin{enumerate}
    \item All bands entirely bluewards of the Lyman limit at $\lambda_{\mathrm{rest}}< 912$~\AA\ should be non-detected at the $2\sigma$ level and all bands entirely redwards of the Lyman limit but bluewards of \lya\ should be $<3\sigma$ detected
    \item The first two bands entirely redwards of Lyman-$\alpha$ should be $>8\sigma$ detected, and all other redwards widebands should be $>3\sigma$ detected
    \item The best-fitting EAZY template should be reasonably well fitting $\chi_{\mathrm{red}}^2<3.0$, which is sufficiently better fitting than the low-redshift EAZY run, $\Delta\chi^2=\chi^2_{z_{\mathrm{max}}=4.0}-\chi^2_{z_{\mathrm{free}}}>4.0$
    \item The best-fitting EAZY redshift PDF should be tightly constrained such that $>60\%$ of the PDF lies within $\pm0.1(1+z)$ of the redshift at maximum likelihood
    \item \sextractor\ half light radii, $R_{50}>45\rm mas$ ($1.5~{\mathrm{pix}}$) in every selection band (F277W, F356W, F444W) to remove any potential hot pixels
    \item Unmasked in at least 2 ACS bands and 6 NIRCam bands, which must include F090W, F277W, F356W, F410M, and F444W
\end{enumerate}

It is worth noting that we use the redshift of maximum likelihood as our photometric redshift, and that the wavelength limits of the photometric filters are taken to be the wavelengths at which there is $50\%$ filter transmission. The band ordering surrounding the break used in selection criteria 1 and 2 is complicated by the extended nature of the F850LP filter transmission profile, which overlaps both F814W and F090W; we thus exclude F850LP from use in the sample selection criteria outlined above.

In addition to these criteria, we fit the {\sf sonora bobcat} \citep{Marley2021-SonoraBobcat}, {\sf cholla} \citep{Karalidi2021}, {\sf diamondback} \citep{Morley2024}, {\sf elf owl} \citep{Mukherjee2024} and {\sf LOWZ} \citep{Meisner2021} Y, L, and T-type brown dwarf templates using the \BDFinder\ package\footnote{\url{https://github.com/tHarvey303/BD-Finder}}, removing 18 suspected brown dwarfs with $\chi_{\rm BD, red.}^2\lesssim\chi_{\rm gal, red.}^2$ which appear compact in the JWST/NIRCam cutouts. An improved procedure for brown dwarf removal, including a full proper-motion search, and morphological fitting for size estimates is expected in the {\tt EPOCHS-v2} release paper (Austin et al. in prep.). We remove 10 compact ``little red dots'' (LRDs) with red $\rm F277W - F444W>0.7$ colours \citep[see e.g.][for photometric LRD selection]{Greene2024, Kokorev2024} from the sample, and an additional 18 spurious candidates are removed by eye, which largely fall within stellar diffraction spikes missed by our automated masking procedure. This leaves us with a final sample consisting of $1721$ $5.6<z<6.5$ galaxies across $549.6$\arcmin$^2$ of unmasked sky area.

\subsection{Photometric properties}

\subsubsection{Calculating UV properties}
\label{sec:UV_properties}

The rest-frame UV properties, including UV continuum slopes, $\beta_{\rm UV}$, and absolute magnitudes, \MUV, for our $5.6<z<6.5$ galaxy sample are calculated directly from the rest-frame UV photometry rather than the best-fitting SED from either \eazypy\ or \bagpipes\ in the same way as \citet{Austin2024}. While this makes only a minor difference for measurements of \MUV, the SED fitting $\beta_{\rm UV}$ measurements are strongly impacted by the parameter space covered by the SED template set being fitted, excluding the extreme blue $\beta_{\rm UV}$ results seen at high redshift \citep[e.g.][]{Cullen2023, Topping2024a, Austin2024, Cullen2024}.

To calculate $\beta_{\rm UV}$ for our sample, we adopt the methodology from \citet{Austin2024}, which fits a $f_{\lambda}\propto\lambda^{\beta_{\rm UV}}$ power law to the photometric fluxes lying entirely within $1250<\lambda_{\mathrm{rest}}/$\AA\ $<3000$, as determined by wavelengths corresponding to their upper and lower $50\%$ filter throughput. We measure \MUV\ from these power law fits using a $100$\AA\,-wide tophat filter centered at $\lambda_{\mathrm{rest}}=1500$\AA.

Since our $\beta_{\rm UV}$ measurements are used only for the calculation of \fesc, we do not correct our $\beta_{\rm UV}$ measurements for the biases discussed in \citet{Austin2024}. While these biases may have an impact on the shape of our $\beta_{\rm UV}-M_{\rm UV}$ relation, the $8\sigma$ detection limit and availability of the F162M, F182M, and F210M medium band filters in a subset of our sample will limit this.


\subsubsection{SED fitting with Bagpipes}
\label{sec:Halpha_bagpipes}

We perform Bayesian SED fitting with a modified version of \bagpipes\ v1.2.0 \citep{Carnall2018}\footnote{This now includes a prescription for \xiionzero\ and \Ndotionzero, among other more minor updates, and is available at \url{https://github.com/tHarvey303/bagpipes}.} adopting a Gaussian redshift prior centred on the ``fsps\_larson'' \eazypy\ redshift with width corresponding to $3\sigma_{\rm z, EAZY}$, where $\sigma_{\rm z, EAZY}$ is the standard deviation of the \eazypy\ redshift PDF. We use \bpass\ v2.2.1 stellar population synthesis models \citep{Eldridge2017-BPASS, Stanway2018-BPASS} with an assumed \citet{Kroupa2001} IMF and a standard \citet{Calzetti2000} dust attenuation curve with no additional dust contribution from birth clouds. The ``continuity bursty'' non-parametric star formation history (SFH) 6-bin prescription from \citet{Leja2019, Tacchella2022}, with an additional low-age bin from $0-3~\rm Myr$ and $3-10~\rm Myr$ is included. We follow the prescription of \citet{Inoue2014} for the attenuation of Lyman-series photons by the IGM. Log-linear priors are set for the dust attenuation, $A_V\in[10^{-4}, 10^2]$, and metallicity, $Z\in[10^{-3}, 3]$, and a flat uniform prior on the logarithm of the ionization parameter ($-4<\log U<0$) is assumed. The upper limit of this $\log U$ range is expanded relative to the default \bagpipes\ BC03 setup informed by \cloudy\ modelling \citep{Ferland2017-CLOUDY}. 

Since the aperture fluxes are input into \bagpipes, the output \MUV\ and $M_{\star}$ properties require an extended source correction. This factor is computed as the ratio of the \sextractor\ ``AUTO\_FLUX'' to ``APER\_FLUX'' in the reddest F444W band, and is clipped to the range $(1, 10)$. This prevents over-correction in cases where \sextractor\ produces catastrophic \citet{Kron1980} aperture sizes, while also ensuring that fluxes for compact, point-like sources are not reduced.

\subsection{Spectroscopic comparison}
\label{sec:EPOCHS_v2_spec_comparison}

To determine the robustness of our photometric redshifts and sample selection procedure, we compile a PRISM/CLEAR spectroscopic catalogue from the most up-to-date version (v4.2) of the Dawn JWST Archive (DJA)\footnote{\url{https://dawn-cph.github.io/dja/}} by cross-matching our photometric catalogue to sources with ``grade 3'' (i.e. the highest quality) with a maximum $0.3$\arcsec\ separation. Several programs comprise our spectroscopic dataset, including from JADES (PI: N. L\"{u}etzgendorf; PIDs: GTOs 1210, 1212, 1286 (GOODS-South), 1214 (COSMOS), 1215 (UDS) + PI: K. Isaak; PID: GTO 1211 \citep[GOODS-North;][]{Maseda2024} + PI: D. Eisenstein; PIDs: GTO 1180 \citep[GOODS-South;][]{D'Eugenio2025}, 1181 (GOODS-North), GO 3215 \citep{Eisenstein2023}), UDS CAPERS (PID: GO 6368, PI: M. Dickinson) and RUBIES \citep[PID: GO 4233; PI: A. de Graaff;][]{deGraaff2024}, COSMOS DDT 6585 (PI: D. Coulter), GO 2565 \citep[PI: K. Glazebrook;][]{Nanayakkara2025} in COSMOS and UDS, as well as DDT 6541 (PI: E. Egami) and GO 2198 \citep[PI: L. Barrufet;][]{Barrufet2025} in GOODS-South. No spectroscopic observations with JWST/NIRSpec are available in the NEP field. 

\begin{figure}
    \centering
    \includegraphics[width=\linewidth]{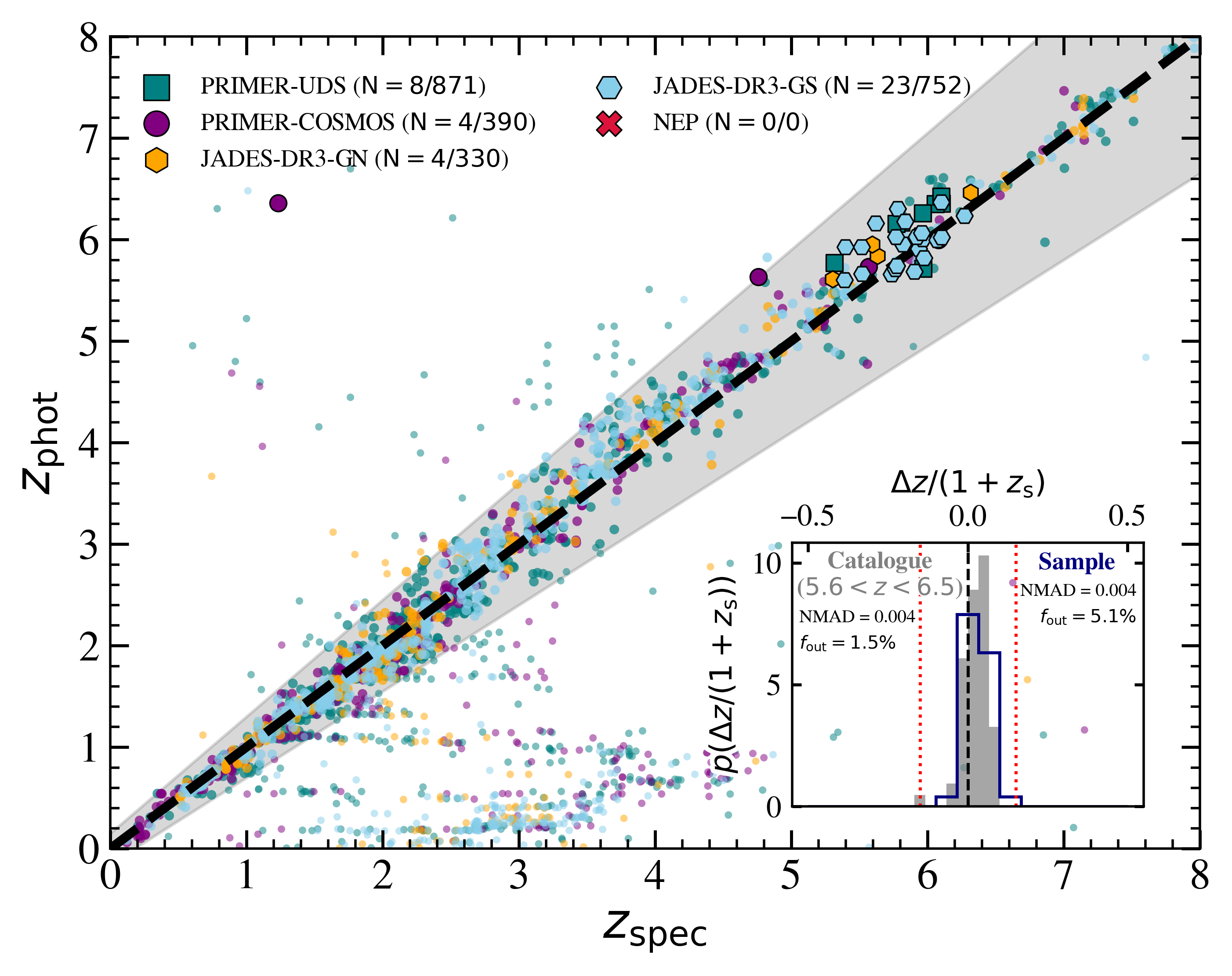}
    \caption[Comparison of \eazypy\ photo-z's with spec-z's collated from the DJA (v4.2)]{Comparison of \eazypy\ photo-z's derived from the \citet{Larson2023} template set with spec-z's collated from the DJA (v4.2). Colours indicate the photometric origin survey, where the larger markers show galaxies that are selected in the photometric sample, of which there are 39. The lower right inset panel shows the probability distribution in $\Delta z/(1+z_s)$ for the catalogue (shaded grey) and sample (navy blue outline), with NMAD and outlier fraction (defined as $\left|\Delta z/(1+z_s)\right| > 0.15$; dotted red lines in inset, and shaded grey in main plot) displayed. The $z_p=z_s$ line is shown in dashed black.}
    \label{fig:EPOCHS_v2_specz_comparison}
\end{figure}

A comparison of photometric and spectroscopic redshifts is presented in \autoref{fig:EPOCHS_v2_specz_comparison}. Our full catalogue has a total 2725 cross-matches across the entire redshift range, with 1169 in UDS, 427 in COSMOS, 355 in GOODS-North, and 774 in GOODS-South. Of these, 39 make our $5.6<z<6.5$ selection criteria as outlined in \autoref{sec:EPOCHS_v2_sample_selection} (8 in UDS, 4 in COSMOS, 4 in GOODS-North, and 23 in GOODS-South). There are 102 PRISM/CLEAR spectra in the full catalogue at $5.6<z<6.5$, although after cropping to just those sources with rest-frame UV $\mathrm{SNR}_{\rm F115W}>\{8, 10\}$, 44 and 39 spectra are retained respectively. This implies a $\gtrsim90\%$ completeness level for our sample which is primarily selected with an $8.0\sigma$ detection limit redwards of \lya, however this is strongly dependent on the selection function for the highly incomplete spectroscopic sample, and hence this number should perhaps be loosely interpreted. A more thorough discussion regarding the completeness of our sample is presented in \autoref{sec:EPOCHS_v2_sample_completeness}.

We quantify the robustness of our photometric redshift estimates by computing both the normalized median absolute deviation (NMAD; $\mathrm{NMAD}=1.483\times\mathrm{median}(\left|\Delta z / (1+z_{\rm s})\right|)$, where $\Delta z = z_{\rm p} - z_{\rm s}$) and the outlier fraction, $f_{\rm out}$, defined as the fraction of galaxies with $\left|\Delta z/(1+z_{\rm s})\right|>0.15$. As can be seen in \autoref{fig:EPOCHS_v2_specz_comparison}, our sample has $f_{\rm out}=5.1\%$ and $\mathrm{NMAD}=0.003$. We note that just a small change in the photo-z posterior for the COSMOS source at $(z_{\rm s}, z_{\rm p}) \simeq (4.8, 5.6)$ towards lower photo-z would halve our $f_{\rm out}$ value as a result of our small sample size. 

We find $f_{\rm out}=28.3\%$ and $\mathrm{NMAD}=0.016$ for our catalogue cropped to $\mathrm{SNR}_{\rm F444W}>8.0$, and $f_{\rm out}=1.5\%$ and $\mathrm{NMAD}=0.004$ at $5.6<z_{\rm s}<6.5$. These photo-z estimates are clearly better suited to higher redshift ($z\gtrsim5$) studies. Improving lower-redshift estimates would require additional bluer filters either from \hst\ or ground-based optical instruments such as Canada-France-Hawaii Telescope's (CFHT) MegaCam \citep{Boulade2003} or Subaru's HyperSuprimeCam \citep[HSC;][]{Miyazaki2018, Komiyama2018}.


\subsection{Bursty and mini-quenched subsamples}

\begin{figure*}
    \centering
        \begin{subfigure}[b]{0.495\textwidth}
        \centering
            \includegraphics[width=\linewidth]{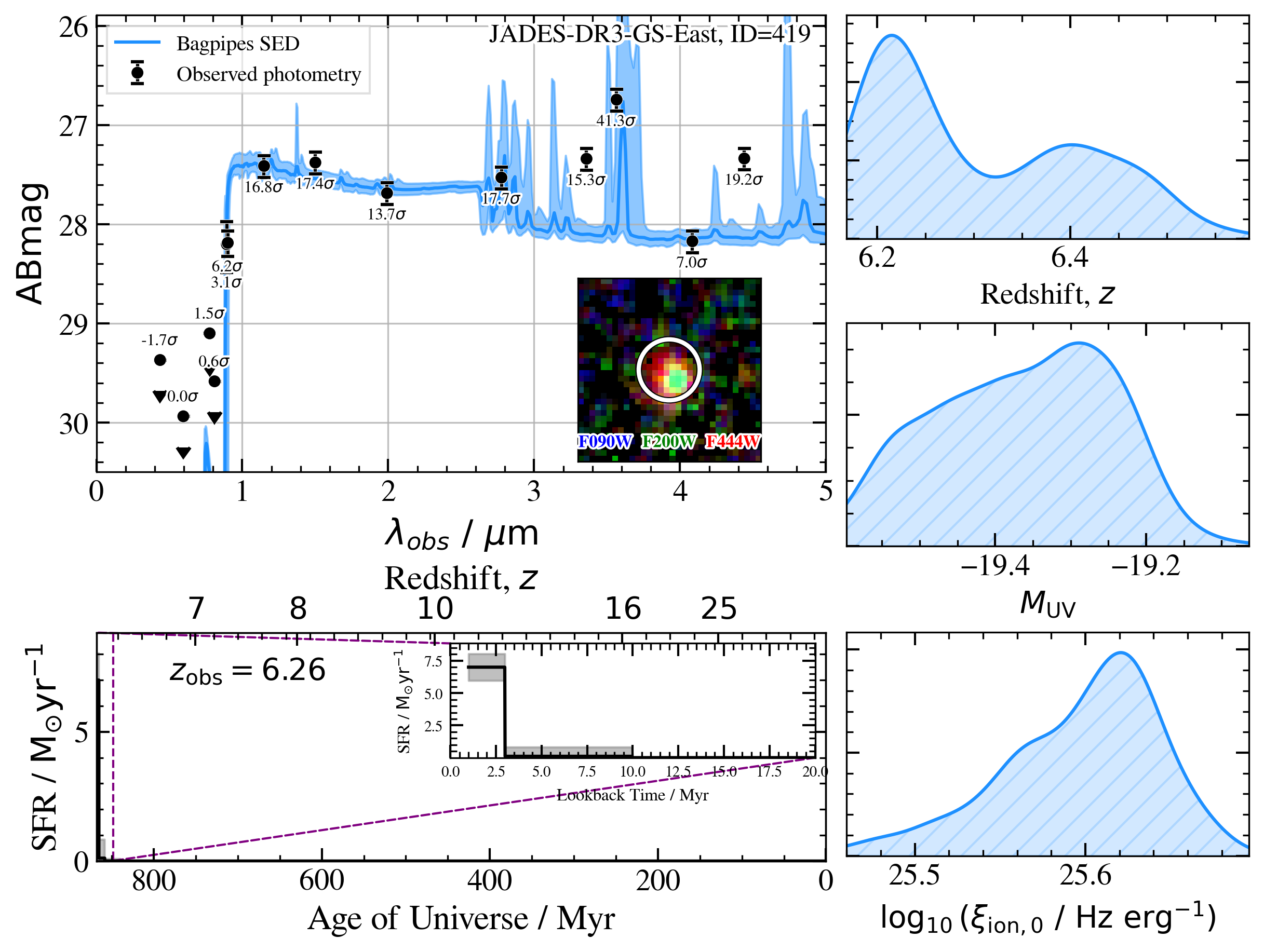}
        \end{subfigure}
        \begin{subfigure}[b]{0.495\textwidth}
        \centering
            \includegraphics[width=\linewidth]{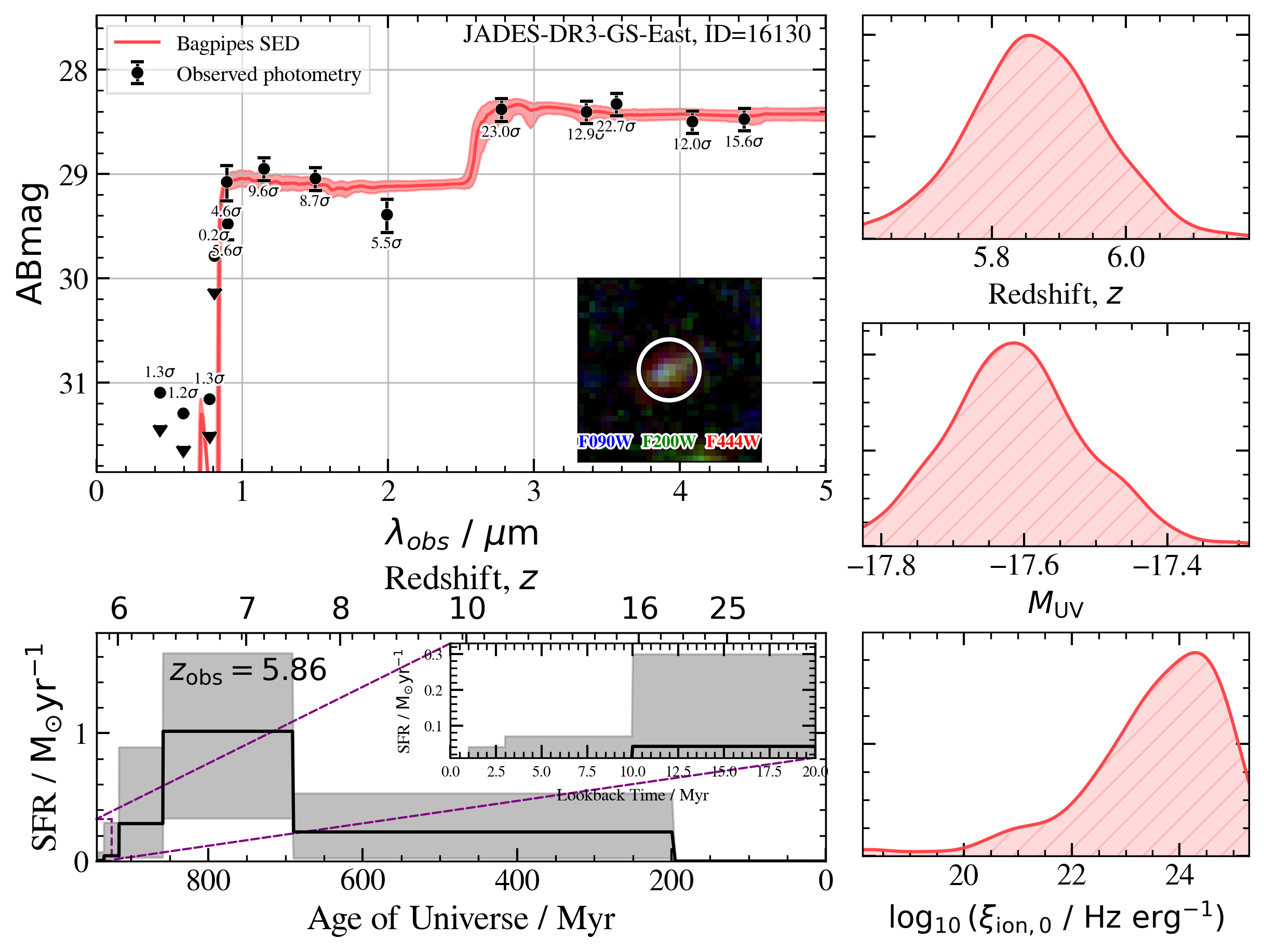}
        \end{subfigure}
    \caption{Example SED fits for star forming (\textbf{left figure}, light blue) and smouldering galaxies (\textbf{right figure}, light red) in the JADES GOODS-South (within our ``East'' imaging) using the fiducial \bagpipes\ setup as outlined in \autoref{sec:Halpha_bagpipes}. In the main panel of each figure, we show as the best-fit SED with shaded $16^{\rm th}-84^{\rm th}$ percentiles. An F090W/F200W/F444W false-colour cutout is shown in the lower right. The photometric magnitudes and SNRs are shown for each observed band; fluxes with $\rm SNR<2$ are shown as upper limits. Posterior redshift, \MUV, and \xiionzero\ PDFs are shown in the right hand panels and the star formation histories are shown in the lower left.}
    \label{fig:sed_plots}
\end{figure*}

We divide our sample into both ``star forming'' and ``smouldering'' subsamples to analyze the contribution of each to reionization independently by the inferred \bagpipes\ star formation burstiness, $\Phi=\mathrm{SFR}_{10}/\mathrm{SFR}_{100}$, such that SFGs have $\Phi>1$ (comprising $62\%$ of the full sample) and vice-versa for smouldering galaxies. In \autoref{fig:sed_plots}, we show example \bagpipes\ SED fitting outputs for star forming (left) and smouldering (right) galaxies taken from the JADES GOODS-South (our ``East'' imaging data) survey. 

The relation between our burstiness and \xiionzero\ estimates is clear from these two systems, where the source in the left-panel with larger \xiionzero\ has a stronger upturn in its recent star formation. Converting the SFRs measured directly from the \ha\ and UV continuum luminosities combined with the conversion factors in \citet{Kennicutt1998} to a burstiness estimate, $\Phi_{\rm H\alpha/\rm UV}=\rm SFR_{\rm H\alpha}/\rm SFR_{\rm UV}$, yields an exact relation with \xiionzero\ given by $\log_{10}(\xi_{\rm ion,0}/\rm Hz~erg^{-1})=25.11+\log_{10}(\Phi_{\rm H\alpha/\rm UV})$.


\subsection{Sample completeness}
\label{sec:EPOCHS_v2_sample_completeness}

We use a single realization of the JWST Extragalactic Mock Catalog \citep[JAGUAR;][]{Williams2018} to assess the completeness of our sample in terms of both \MUV\ and stellar mass, $M_{\star}$. The relevant catalogue fluxes are scattered by $1\sigma$ mock photometric errors determined from the median depth in each band in each survey with a 10\% minimum flux error as per our real photometric catalogues. For the JADES-GS survey we repeat this procedure for the deeper JADES-DR1 survey area and JOF which contains deep medium band imaging. Since the PRIMER-UDS and PRIMER-COSMOS fields have tiered NIRCam exposure times we divide these surveys into Deep and Wide regions, producing a separate catalogue for each.

\begin{figure}
    \centering
        \begin{subfigure}[b]{0.48\textwidth}
        \centering
            \includegraphics[width=\linewidth]{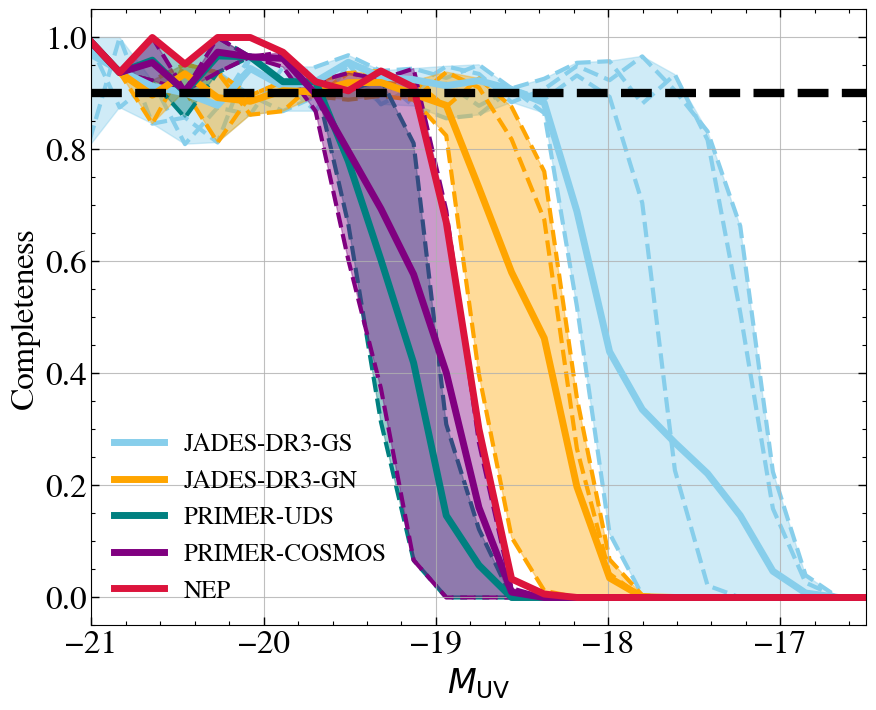}
        \end{subfigure}
        \hfill
        \hfill
        \begin{subfigure}[b]{0.48\textwidth}
        \centering
            \includegraphics[width=\linewidth]{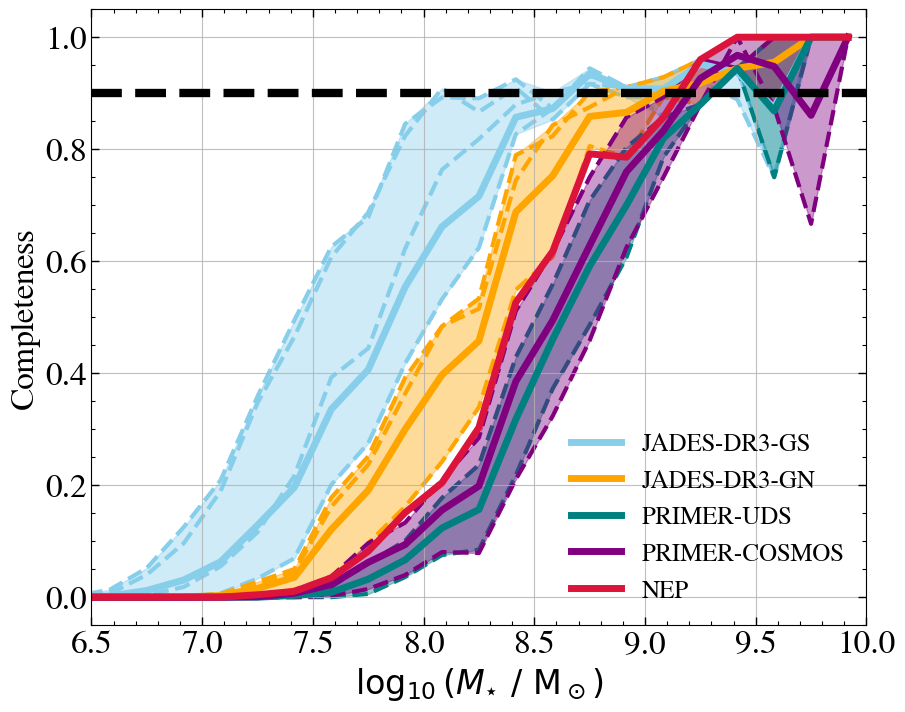}
        \end{subfigure}
    \caption[JAGUAR Sample completeness curves in terms of \MUV\ and $M_{\star}$ coloured by EPOCHS v2 survey]{Sample completeness curves from our mock JAGUAR catalogues \citep{Williams2018} coloured by EPOCHS v2 survey. Solid coloured lines show the average completeness of the combined catalogue for a particular survey and the thinner dashed lines show the results for each individual catalogue. The thick, dashed, black horizontal line shows the 90\% completeness limit. We note that the brightest and most massive bins have a reduced sample size, and more JAGUAR realizations are required to average over this variance. \textbf{Upper panel)} Completeness defined in terms of \MUV. \textbf{Lower panel)} Completeness curves defined in terms of input \beagle\ stellar mass, $M_{\star}$.}
    \label{fig:EPOCHS_v2_sample_completeness}
\end{figure}

We run our catalogues through \eazypy\ and the following selection procedure (excluding the masking criteria) outlined in \autoref{sec:EPOCHS_v2_sample_selection}, with \MUV\ and $M_{\star}$ completeness curves for each survey shown in \autoref{fig:EPOCHS_v2_sample_completeness}. It is noteworthy that we do not determine stellar masses using our \bagpipes\ setup due to computational limitations; recovered \bpass\ stellar masses may show a systematic difference compared to the {\tt JAGUAR} masses which are determined from \beagle\ templates as a result of a differing assumed IMFs ({\tt JAGUAR} assumes \citet{Chabrier2003} compared to the \bagpipes\ \citet{Kroupa2001} IMF), SFHs, or assumed dust content. 

Assuming that the stellar masses are robustly recovered, we find that our photometric sample is 90\% mass complete between $\log_{10}(M_{\star}/\rm M_{\odot})=8.1$ (JADES-DR1 and JOF) and $\log_{10}(M_{\star}/\rm M_{\odot})=9.3$ (PRIMER-UDS Wide). The more step-function like \MUV\ completeness curves show that we are 90\% complete between $M_{\rm UV}=-17.6$ (JADES-DR1 and JOF) and $M_{\rm UV}=-19.7$ (PRIMER-UDS and PRIMER-COSMOS Wide regions). All scaling relations in this work (presented in \autoref{sec:EPOCHS_v2_xi_ion_MUV}, \autoref{sec:EPOCHS_v2_ndot_ion_MUV}, and \autoref{sec:EPOCHS_v2_beta_MUV}) are computed with a 90\% stellar mass complete sample as in \citet{Simmonds2024b} to ensure the relations remain unbiased from faint blue galaxies bursting into the selected sample.

\section{The ionizing photon production efficiency}
\label{sec:xi_ion}

\subsection{Calculating \texorpdfstring{$\xi_{\rm ion}$}{ionizing photon production efficiencies}}
\label{sec:calculating_xi_ion}


To determine the contribution of our galaxy sample to reionization, it is useful to compute both the ionizing photon production rate, 
\begin{equation}
    \dot{N}_{\mathrm{ion}} = \frac{7.28\times10^{11}}{(1-f_{\rm esc}^{\rm LyC})}L_{{\mathrm{H}\alpha}} \quad [\rm Hz],
    \label{eq:Ndot_ion}
\end{equation}
and efficiency,
\begin{equation}
   \xi_{\mathrm{ion}} = \frac{\dot{N}_{\mathrm{ion}}}{L_{\nu, \rm UV}} \quad [\rm Hz~erg^{-1}], 
   \label{eq:xi_ion_definition}
\end{equation}
where $L_{{\mathrm{H}\alpha}}$ is the \ha\ luminosity and $L_{\nu, \rm UV}$ is the UV continuum luminosity density, which are both dust corrected. The conversion factor in \autoref{eq:Ndot_ion} assumes case B recombination and a standard nebular electron temperature and density \citep[see][]{Osterbrock2006}. 
These properties, however, are not independent of \fesc\ and it is more common to instead compute \Ndotion\ and \xiion\ whilst assuming \fesc$=0$, such that
\begin{equation}
    \dot{N}_{\rm ion,0}=(1-f_{\rm esc}^{\rm LyC})\dot{N}_{\rm ion}; \quad \xi_{\rm ion,0}=(1-f_{\rm esc}^{\rm LyC})\xi_{\rm ion}.
    \label{eq:escaping_ndot_xi_ion}
\end{equation}

Our modified \bagpipes\ calculates both \xiionzero\ and \Ndotionzero\ for each source from the full posterior fitting distributions, assuming $f_{\rm esc}^{\rm LyC}=0$. To calculate the rates/efficiencies of escaping ionizing photon production, we must multiply these observed (\fesc\ independent, but dust corrected) values output by \bagpipes\ by a factor $f_{\rm esc}^{\rm LyC}/(1-f_{\rm esc}^{\rm LyC})$. In addition, we note that we also perform extended source corrections for \Ndotionzero\ in the same manner as the stellar mass correction factors explained in \autoref{sec:Halpha_bagpipes}. This \Ndotionzero\ is not to be confused with the \textit{total} rate of ionizing photon production, \ndotion, introduced in \autoref{sec:total_reionization_contribution} and given by \autoref{eq:total_ndot_ion}.



%
%
%

\subsection{Robustness of our \bagpipes\ \texorpdfstring{\xiionzero}{ionizing photon production efficiency} results}
\label{sec:xi_ion_spec}

\begin{figure*}
    \begin{subfigure}[b]{0.5\textwidth}
        \includegraphics[width=\textwidth]{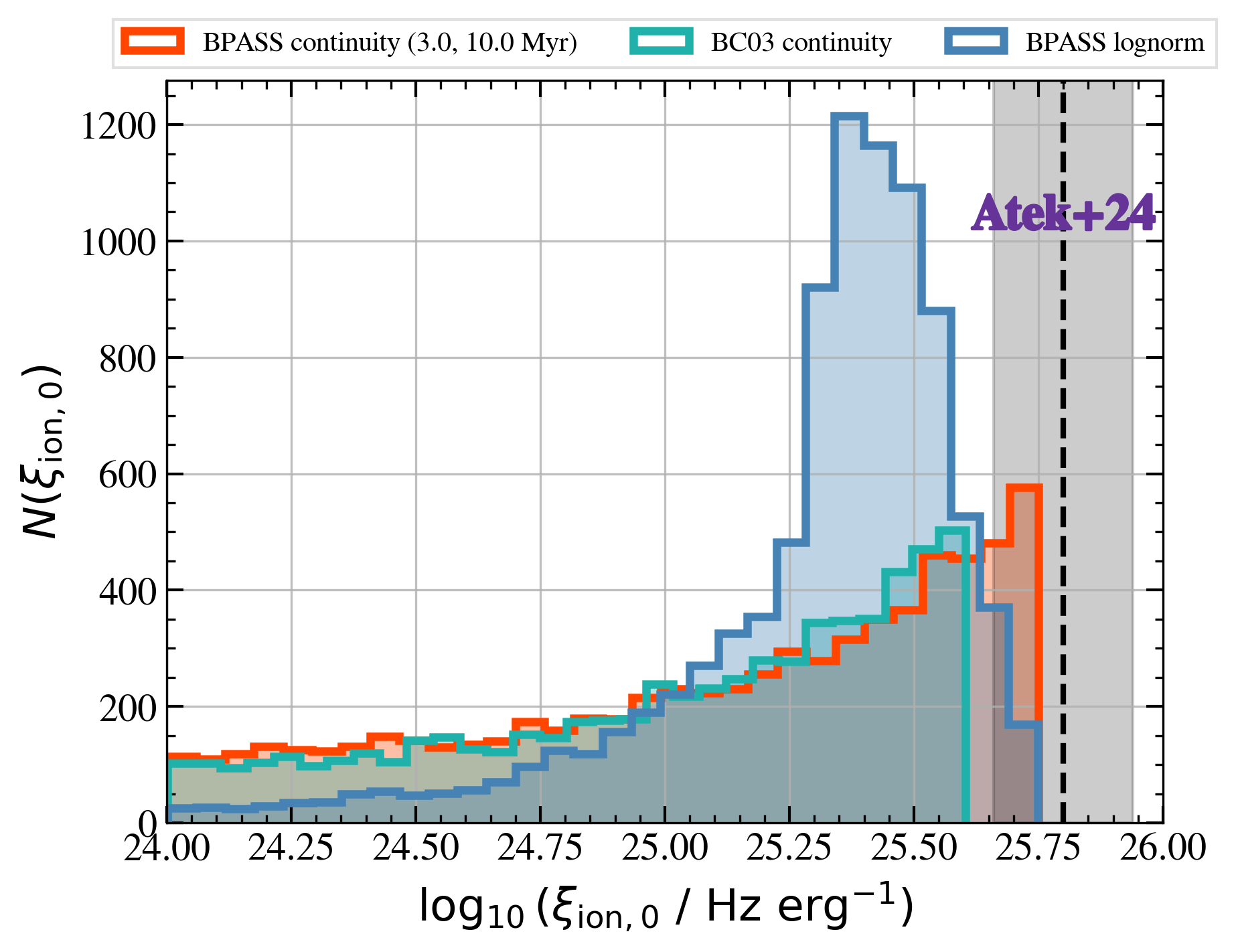}
    \end{subfigure}
    \begin{subfigure}[b]{0.5\textwidth}
        \includegraphics[width=\textwidth]{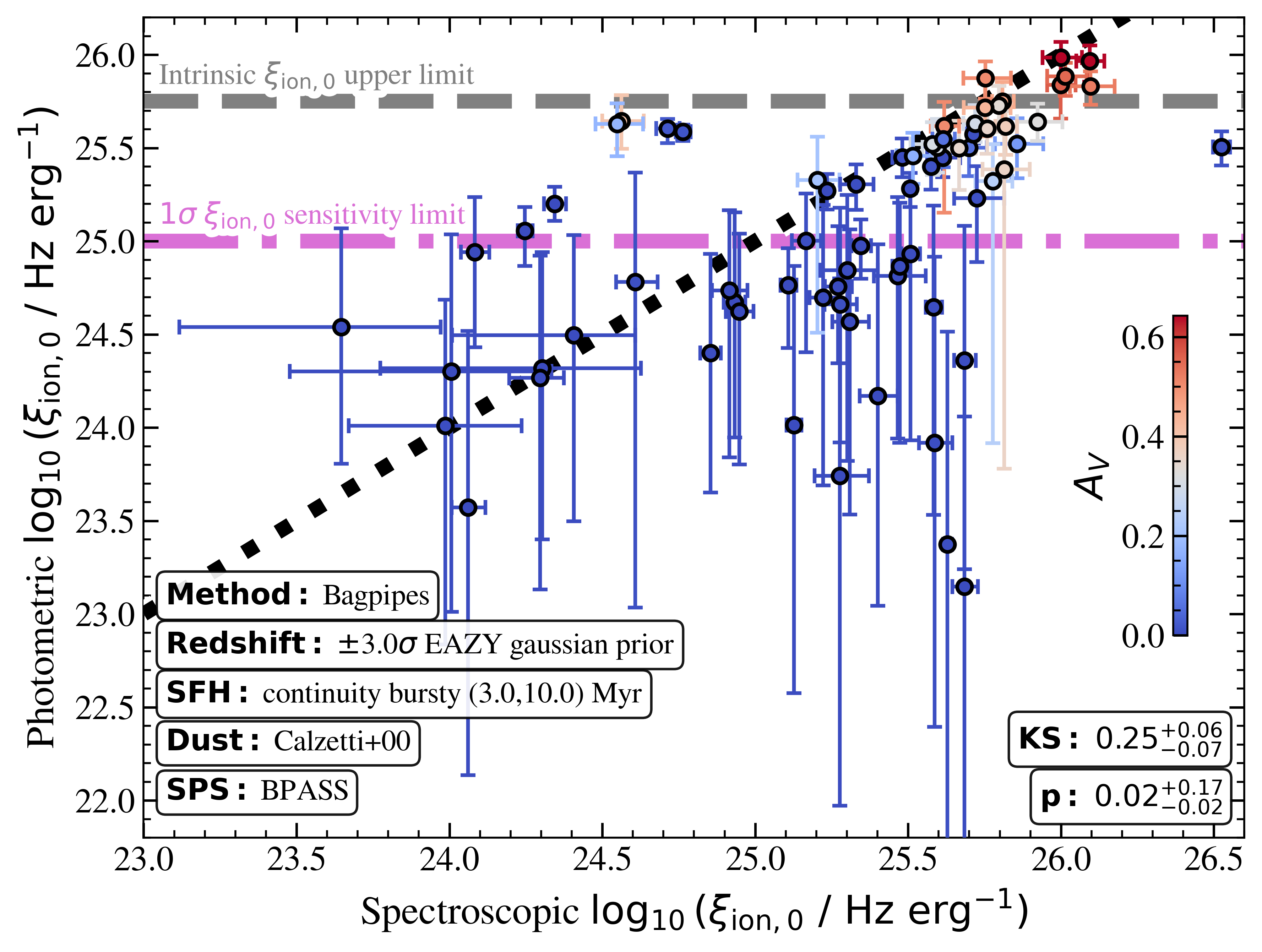}
    \end{subfigure}
\caption[\bagpipes\ \xiionzero\ priors for a number of different {\tt BC03} and \bpass\ setups and spectroscopic comparison of photometrically derived $\xi_{\rm ion}$]{\textbf{Left panel)} \bagpipes\ \xiionzero\ priors for our fiducial setup (orange red), with {\tt BC03} and the standard \citet{Tacchella2022} ``continuity bursty'' SFH (sea green) and \bpass\ lognormal SFH (blue) also shown. Mean \xiionzero\ results from UNCOVER by \citet{Atek2024} are shown for comparison. \textbf{Right panel)} Comparison of spectroscopic \textit{observed} \xiionzero\ measured from narrow \ha\ fits to 71 $5.6<z<6.5$ NIRSpec PRISM/CLEAR spectra from the DJA against mock photometric measurements computed using our fiducial \bagpipes\ setup. The intrinsic (i.e. dust free) \xiionzero\ upper prior limit from the left-hand panel is shown as a dashed grey line and the $1\sigma$ \xiionzero\ sensitivity limit at $\log_{10}(\xi_{\rm ion,0}/\rm Hz~erg^{-1})=25.0$ is shown in purple. We colour the \xiionzero\ measurements by the photometrically determined V-band dust attenuation, $A_V$. The result of a bootstrapped KS goodness-of-fit test and corresponding $p$-value is shown in the lower right.}
\label{fig:bagpipes_xi_ion}
\end{figure*}

We test the priors for our fiducial \bagpipes\ setup explained above, finding that the possible values for $\xi_{\rm ion, 0}$ strongly depend upon both the SPS model used and the SFH. In the left-hand panel of \autoref{fig:bagpipes_xi_ion} we show the \bagpipes\ priors for the setup used in this work (orange red) and compare to the priors from two additional setups. The first alternative setup assumes a {\tt BC03} SPS and ``continuity bursty'' SFH without the low-age bins (sea green), whereas the second adopts a \bpass\ SPS with lognormal SFH (blue). All other fiducial parameters remain fixed. It is apparent that while the stellar emission from our fiducial \bagpipes\ setup does the best, the intrinsic values still cannot match the large $\langle\log_{10}(\xi_{\rm ion}/\rm Hz~erg^{-1})\rangle=25.8\pm0.14$ measured for lensed galaxies in the UNCOVER survey at $6<z<8$ with $-17<M_{\rm UV}<-15$ by \citet{Atek2024} due to the limited parameter space explored by the stellar template set run through \cloudy.

To determine the robustness of our individual \xiion\ measurements, we produce a secondary spectroscopic catalogue from v4.2 of the DJA, finding 663 ``grade 3'' PRISM/CLEAR spectra at $5.6<z<6.5$. We perform continuum SNR cuts at the rest-frame \ha\ wavelength at $2\sigma$ as well as a $5\sigma$ cut at $\lambda_{\rm rest}=1500$~\AA, leaving 83 remaining spectra, 8 of which fail the selection criteria outlined in \autoref{sec:EPOCHS_v2_sample_selection}. The \ha\ lines in these 75 spectra were fit with a Gaussian profile using the {\tt specFitMSA}\footnote{\url{https://github.com/vadimrusakov/specFitMSA}} python package. UV magnitudes were calculated using the same tophat filter as for our photometry to determine \textit{observed} (i.e. not dust corrected) $\xi_{\rm ion}$ values spectroscopically. 

A comparison of spectroscopic and photometric \textit{observed} \xiionzero\ measurements are shown in the right-hand panel of \autoref{fig:bagpipes_xi_ion}, where we additionally remove 4 sources with broad \ha. The intrinsic \xiionzero\ upper prior limit is plotted as the horizontal dashed grey line; this comparison plot shows that when \bagpipes\ can match the observed \ha\ fluxes with a dust-free stellar population it does so, and in the case of $\xi_{\rm ion, obs}>\xi_{\rm ion, intr}$, \bagpipes\ instead assigns increasing amounts of dust until the F444W \ha\ flux is reached. This will bestow large $A_V$ (as much as $\sim0.6~\rm mag$) values to the youngest, most highly star forming systems, which are likely dust-free due to the timescales required for asymptotic giant branch (AGB) stars to produce copious amounts of dust.

As explored in \autoref{sec:xi_ion_sensitivity}, there also exists a limiting sensitivity to which we can detect a flux boost in F444W compared to the continuum as measured in F410M. Below this sensitivity limit ($\log_{10}(\xi_{\rm ion,0}/\rm Hz~erg^{-1})=25.0$ at $1\sigma$), the SED fitting becomes biased towards lower photometric \xiionzero. A bootstrapped Kolmogorov-Smirnov (KS) goodness-of-fit test is performed over $100,000$ steps, from which we measure $\mathrm{KS}=0.25^{+0.06}_{-0.07}$ and $p=0.02^{+0.17}_{-0.02}$. This low $p$-value suggests that systematics do indeed exist in the SED fitting process, in part due to the prior limit imposed by limitations of modeling the youngest stars and the \xiionzero\ sensitivity limit.

\subsection{Ionizing photon production efficiency as a function of \texorpdfstring{$M_{\rm UV}$}{intrinsic UV magnitude}}
\label{sec:EPOCHS_v2_xi_ion_MUV}

\begin{figure}
    \centering
    \includegraphics[width=1.05\linewidth]{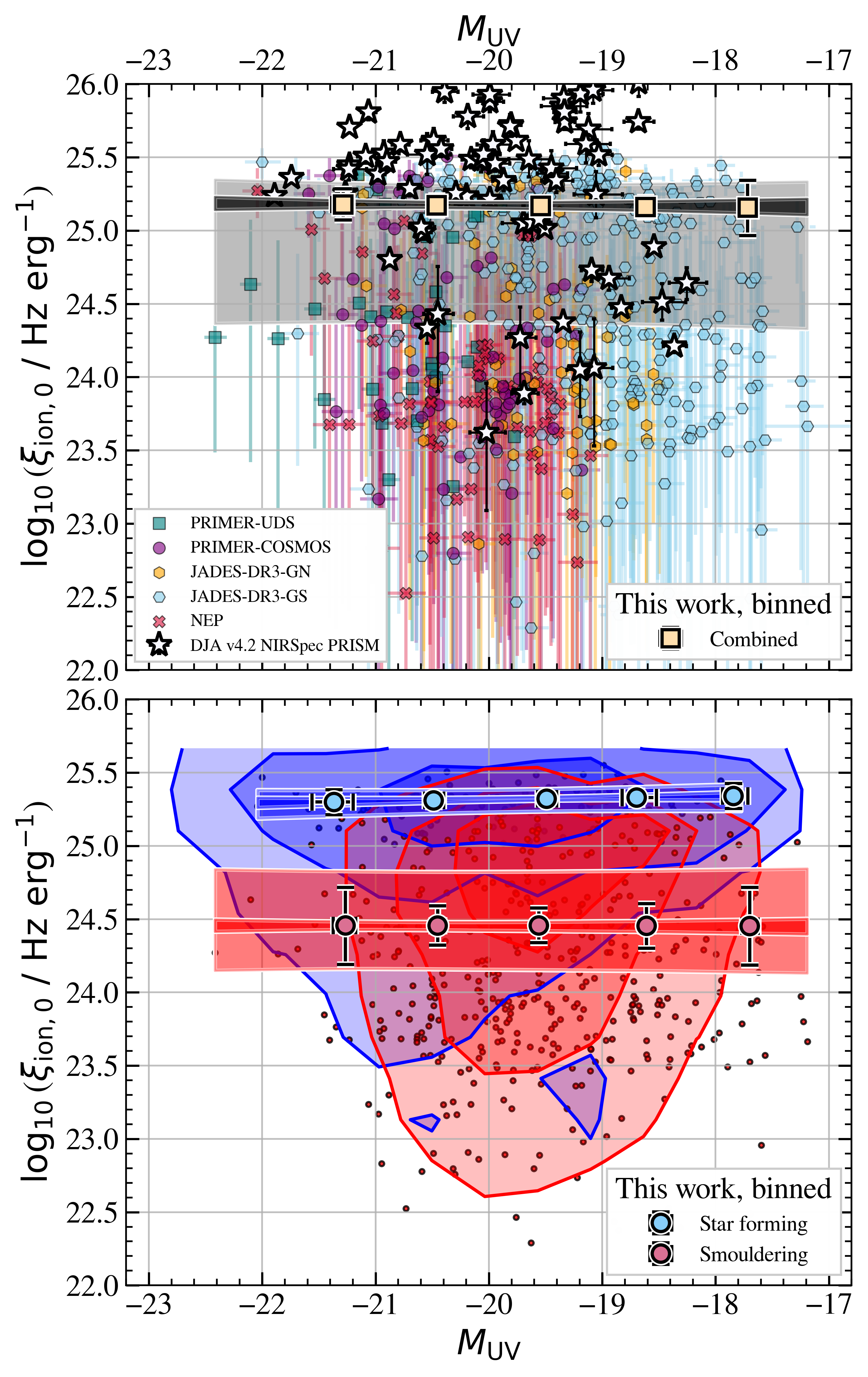}
    \caption{$\xi_{\rm ion,0}-M_{\rm UV}$ derived from our 90\% stellar mass complete sample. Darker shaded regions show the $16^{\rm th}-84^{\rm th}$ percentiles of the linear fit, with lighter shaded regions showing the scatter in the data. Fitting parameters are given in \autoref{tab:power_law_params}. Wild bootstrap-binned data points calculated using \autoref{eq:wild_bootstrapping} are displayed in $1~\rm dex$ \MUV\ bins. \textbf{Upper panel)} Results are coloured by photometric origin field, with white stars showing measurements from our 71 selected NIRSpec/PRISM spectra. \textbf{Lower panel)} Results obtained for star forming (blue) and smouldering (red) subsamples. The respective contours show $\{1,2,3\}\sigma$ confidence intervals.}
    \label{fig:EPOCHS_v2_xi_ion_MUV}
\end{figure}

The \MUV\ dependence of our \xiionzero\ estimates for our 90\% stellar mass complete sample is presented in \autoref{fig:EPOCHS_v2_xi_ion_MUV}. In the upper panel we split by survey origin; the bright end is dominated by our widest PRIMER-UDS and PRIMER-COSMOS surveys and the faint end is dominated by the deepest JADES-DR3-GS data. White stars correspond to measurements from the DJA v4.2 spectroscopic sample of 71 sources outlined in \autoref{sec:xi_ion_spec}. These spectroscopic \xiionzero\ values are dust-corrected with $A_{\rm UV}$ determined via conversion from $\beta_{\rm UV}$ (measured directly from the PRISM spectra in the 10 \citet{Calzetti1994} filters) assuming the \citet{Meurer1999} conversion factor, and hence show characteristically larger values than in the right-hand panel of \autoref{fig:bagpipes_xi_ion}. The clear trend of \bagpipes\ burstiness with measured \xiionzero\ is clear to see in the lower panel of \autoref{fig:EPOCHS_v2_xi_ion_MUV}, where we plot results from both the star forming and smouldering subsamples.

We utilize Monte Carlo Markov Chain (MCMC) sampling techniques to fit a power law using the {\tt emcee} python package \citep{Foreman-Mackey2013}. Wide uniform priors on the amplitude and slope are adopted with limits $\pm10\sigma$ from the median value of a preliminary linear least squares fit. Our results suggest that the flat relation
\begin{equation}
    \log_{10}(\xi_{\rm ion, 0}/\rm Hz~erg^{-1}) = -0.006^{+0.019}_{-0.017}~M_{\rm UV} + 25.05^{+0.39}_{-0.34}
    \label{eq:linear_fit_xi_ion_0}
\end{equation}
is most appropriate, consistent with \hst\ canonical values at $M_{\rm UV}\simeq-19$ \citep{Robertson2013}. Best-fit parameters and errors for our SFG and smouldering subsamples are shown in \autoref{tab:power_law_params} in \autoref{sec:power_law_params}. 

Bootstrap-binned data points are additionally shown for the combined, SFG, and smouldering subsamples using the following methodology. First we sample from the \MUV\ and \xiionzero\ PDFs before splitting into evenly spaced $1~\rm dex$ \MUV\ bins. The \xiionzero\ values in each bin are then resampled using the wild bootstrapping scheme
\begin{equation}
    \xi_{\mathrm{ion,0},i}^\prime = f(M_{\mathrm{UV}, i};\hat{\theta}) + \hat{\epsilon}_i\Re,
    \label{eq:wild_bootstrapping}
\end{equation}
for each value, $i$, where $f$ is the power law model evaluated with best-fit parameters $\hat{\theta}$, and $\hat{\epsilon}_i=\xi_{\mathrm{ion,0},i} - f(M_{\mathrm{UV}, i};\hat{\theta})$ is the \xiionzero\ residual to the best-fit model. Since $\sigma_{\xi_{\rm ion,0}}$ scale with \xiionzero, we multiply the residuals by a Rademacher distribution ($\Re$, with probability density $p(k)=\frac{1}{2}\big(\delta(k-1) + \delta(k+1)\big)$) to account for the heteroscedasticity.

\subsection{Scatter in the \texorpdfstring{$\xi_{\rm ion,0}-M_{\rm UV}$}{ionization photon production efficiency vs intrinsic UV magnitude} relation}

From \autoref{fig:EPOCHS_v2_xi_ion_MUV}, it is clear that our measurement errors on \xiionzero\ are non-Gaussian and therefore that our data is heteroscedastic. We thus model the asymmetric scatter in the distribution as $\sigma(\xi_{\rm ion, 0})=\lvert\sigma_{\rm c} + \sigma_{\rm \xi_{\rm ion,0}}~\xi_{\mathrm{ion},0}\rvert$, where the variance on \xiionzero\ in the log-likelihood function is calculated as the quadrature sum of this scatter with the \bagpipes\ measurement uncertainties for each source. Wide uniform priors of $\sigma_{\rm c}\in[10^{-6},100]$ and $\sigma_{\xi_{\rm ion,0}}\in[-100,0]$ are adopted so as to break the bimodality between negative and positive scatter solutions.

The posterior scatter parameters are $\sigma_{\rm c}=18.98^{+1.13}_{-1.07}$ and $\sigma_{\xi_{\rm ion,0}}=-0.754^{+0.043}_{-0.045}$, indicating that the scatter reduces with increasing \xiionzero. These results highlight the uniformity of \xiionzero\ measurements for star-forming galaxies which are currently in a burst phase; smouldering galaxies, which are \textit{on average} older, may have greater diversity in galaxy properties, such as stellar masses or sSFRs. The scatter parameters $\Theta=\{\sigma_c,\sigma_{\xi_{\rm ion,0}}\}$ for our star forming and smouldering subsamples are shown in \autoref{tab:power_law_params}.




\subsection{Literature comparison of \texorpdfstring{\xiionzero$-$\MUV}{the ionization photon production efficiency vs intrinsic UV magnitude}}
\label{sec:xi_ion_MUV_lit_comparison}

We compare our \xiionzero$-$\MUV\ results to the recent literature in \autoref{fig:xi_ion_MUV_lit_comp}. Our results are in contrast with the most recent photometric results, which suggest that \xiionzero\ increases towards brighter UV magnitudes. 

\subsubsection{Photometric studies}

Early JWST-derived results from \citet{Simmonds2024a} suggest a strong positive trend in the opposite direction at $z\simeq6$, mitigated by \citet{Simmonds2024b} who perform a 90\% completeness cut in stellar mass to remove faint low-mass SFGs, which are inherently more detectable than their less bursty counterparts at the same $M_{\star}$, ``star bursting into the sample''. Since \citet{Simmonds2024b} report a 90\% mass completeness limit at $M_{\star}\simeq10^{7.5}~\rm M_{\odot}$, which is $0.5-1.5~\rm dex$ lower than ours in GOODS-North/South (see \autoref{fig:EPOCHS_v2_sample_completeness} in \autoref{sec:EPOCHS_v2_sample_completeness}), we might expect our stellar mass cut to provide a significantly steeper slope, however this is not what is observed. Perhaps the difference stems more from the \prospector\ SED fitting procedure, which isused by \citet{Simmonds2024a} and \citet{Simmonds2024b}, compared to \bagpipes. 

The \citet{Begley2025a} $6.9<z<7.6$ sample derived from \fion{O}{3}+\hbeta\ uses the same \bagpipes\ setup as us, using data from PRIMER and JADES, and report a similar $M_{\star}\simeq10^{7.6}~\rm M_{\odot}$ 90\% mass completeness limit and \xiionzero$-$\MUV\ slope to \citet{Simmonds2024b}. We caution that this result may be influenced by the metallicity dependence of \fion{O}{3}/\hbeta, which cannot easily be constrained by photometry alone, leading to an increase in scatter of scaling relations which convert this quantity to \xiionzero\ \citep[see e.g.][]{Tang2019}. In addition, we note that this \bagpipes\ setup fails to reproduce simultaneously the \fion{O}{3}+\hbeta\ and \ha+\fion{N}{2} fluxes. Nevertheless, the study is comprehensive and it remains challenging to determine the reasons for the difference between our fitted relations.

\subsubsection{Spectroscopic studies}

Spectroscopic studies appear to show wildly differing results, potentially stemming from the unknown and complex Micro-Shutter Assembly (MSA) selection function. In line with our spectroscopic sample at $z\simeq6$, many studies appear to show an increasing \xiionzero\ towards faint \MUV, including those from CEERS, JADES, GLASS, and UNCOVER \citep[][]{Llerena2025a, Papovich2025}, as well as the Early eXtragalactic Continuum and Emission Line
Science \citep[EXCELS; GO 3543; PIs: A. Carnall, F. Cullen;][see \citealt{Begley2025b} for the \xiionzero\ analysis]{Carnall2024} program. Presumably these samples include preferentially bursty, faint galaxies, which have been the source of much intrigue. The work of \citet{Pahl2025a} using CEERS and JADES NIRSpec spectra, and subsequently the Assembly of Ultradeep Observations Revealing Astrophysics (AURORA; PID:1914) sample in \citet{Pahl2025b}, shows an opposing trend. Our photometric observations show a significantly flatter trend than any of these studies would suggest.

\subsubsection{Hydrodynamical simulations and SAMs}

\begin{figure}
    \centering
    \includegraphics[width=0.9\linewidth]{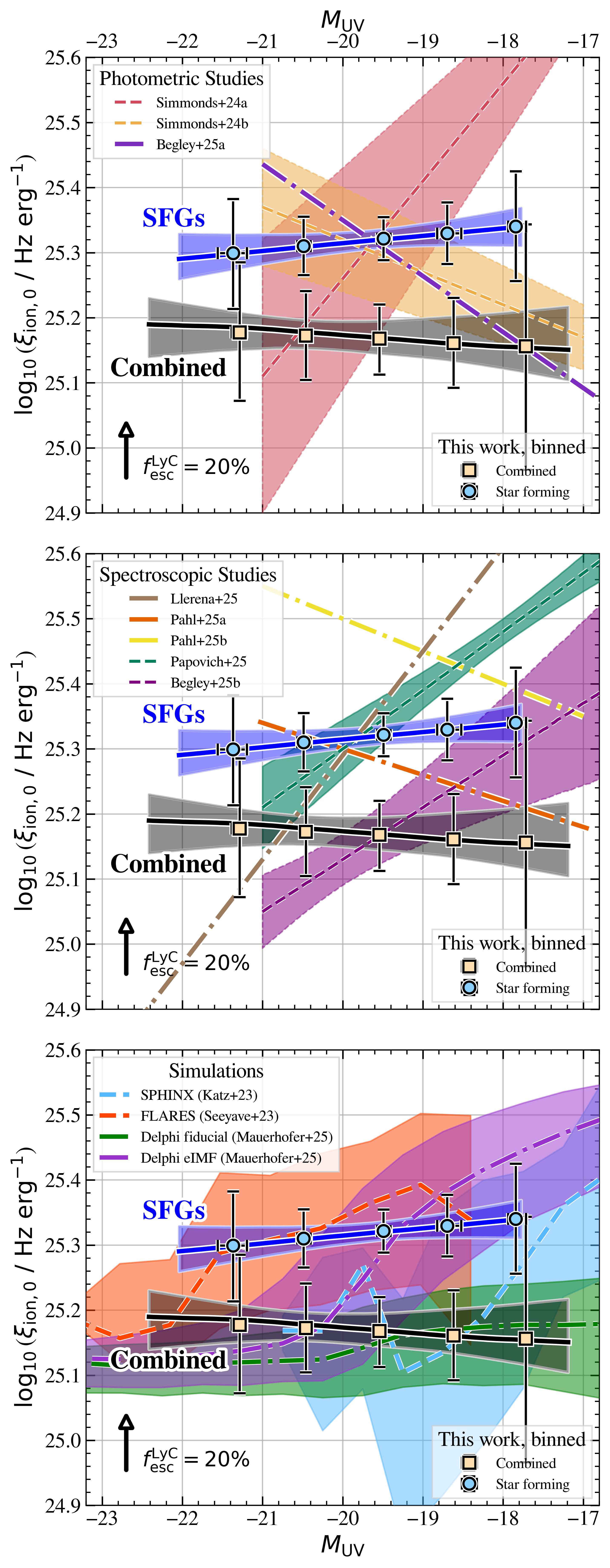}
    \caption{\xiionzero$-$\MUV\ power law fits to our SFG and combined samples, with wild bootstrap bins computed as per \autoref{eq:wild_bootstrapping}. The shaded regions show the $16^{\rm th}-84^{\rm th}$ percentiles. \textbf{Upper panel)} Comparison to photometric studies \citep{Simmonds2024a, Simmonds2024b, Begley2025a}. \textbf{Central panel)} Comparison with spectroscopic studies \citep{Llerena2025a, Pahl2025a, Pahl2025b, Papovich2025, Begley2025b}. \textbf{Lower panel)} Comparison to SPHINX \citep[][light blue]{Katz2023_sphinx}, FLARES \citep[][orange]{Seeyave2023}, and DELPHI \citep[][fiducial in dark green, extended IMF in purple]{Mauerhofer2025} simulations.}
    \label{fig:xi_ion_MUV_lit_comp}
\end{figure}

We additionally compare our results to those from hydrodynamical simulations on large scales from the First Light And Reionization Epoch Simulations \citep[FLARES;][]{Lovell2021_FLARES_I, Vijayan2021_FLARES_II, Wilkins2023_FLARES_V, Seeyave2023}, and SPHINX \citep[][]{Rosdahl2018, Rosdahl2022, Katz2023_sphinx} simulations. Results from the Dark matter and
the Emergence of gaLaxies in the ePocH of reIonization \citep[DELPHI;][]{Dayal2020,Mauerhofer2025} SAM are also shown in \autoref{fig:xi_ion_MUV_lit_comp} for the fiducial (green) and extended IMF (eIMF; light blue) runs. 

Our combined sample is consistent with the fiducial DELPHI run with standard IMF, although our SFG sample is $\sim0.1-0.15~\rm dex$ higher, matching the extended IMF model at \MUV$\gtrsim-20$. This sample best matches the FLARES data at \MUV$\lesssim-20$ and overpredicts SPHINX until \MUV$\simeq-18$. The increase in \xiionzero\ towards faint \MUV\ values in the SPHINX and eIMF DELPHI run is indicative of exotic starburst populations beginning to dominate the sample. This is observed in spectroscopic samples as well as our stellar mass complete photometric sample, although with not quite as extreme an inflection point.

\subsection{\texorpdfstring{\xiionzero$-M_{\rm UV}$}{ionization photon production efficiency vs intrinsic UV magnitude} relation as a function of stellar mass}

\begin{figure}
    \centering
    \includegraphics[width=0.98\linewidth]{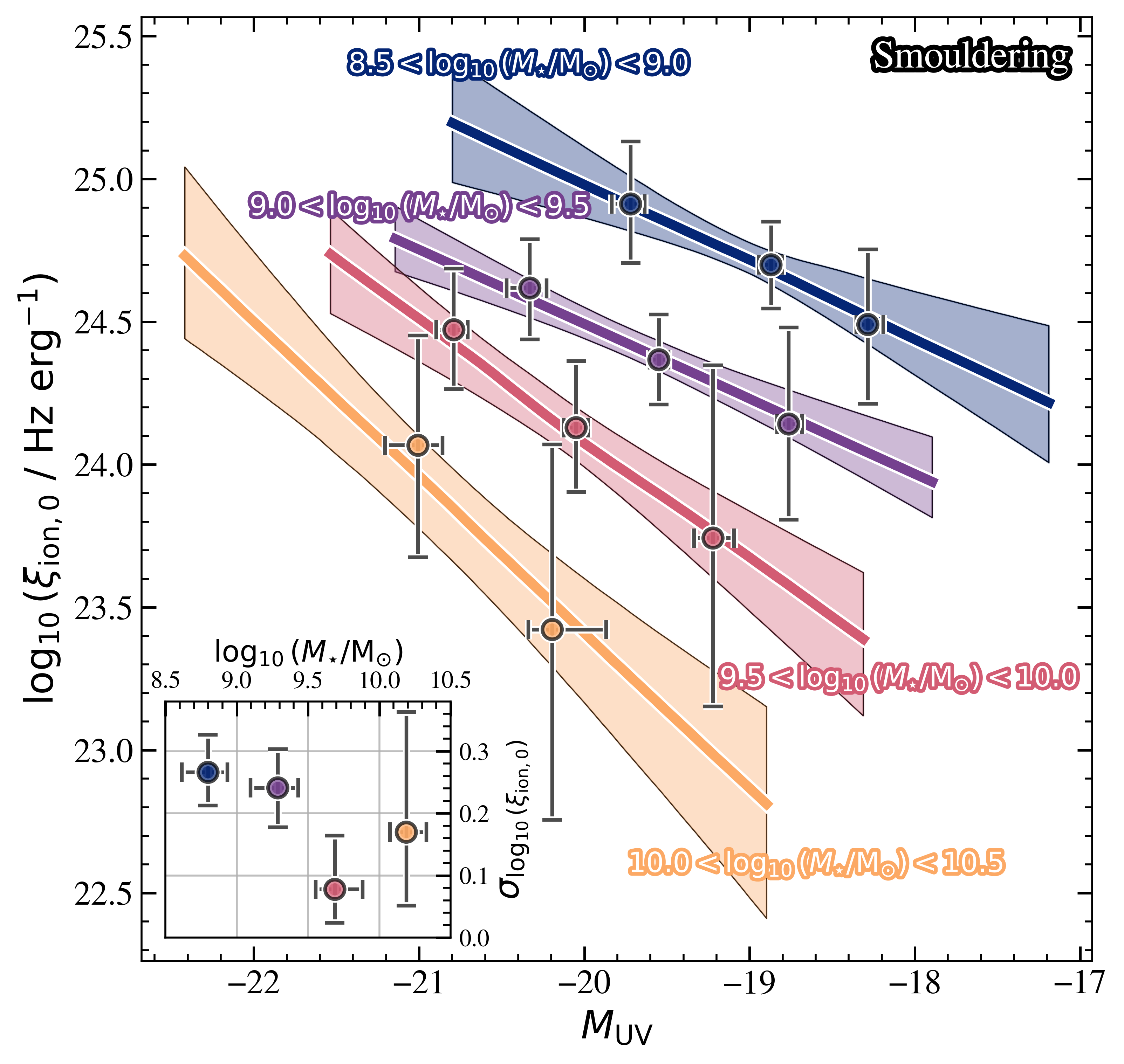}
    \caption{\xiionzero$-$\MUV\ relation for our smouldering subsample in $0.5~\rm dex$ stellar mass bins across the range $8.5<\log_{10}(M_{\star}/\rm M_{\odot})<10.5$. The symmetric scatter, $\sigma_{\log_{10}(\xi_{\rm ion,0})}$, is given as a function of $M_{\star}$ in the lower left inset panel. Wild bootstrapped binned data points are computed as per \autoref{eq:wild_bootstrapping}.}
    \label{fig:xi_ion_MUV_mass_split}
\end{figure}

We fit our \xiionzero$-M_{\rm UV}$ in fiducial \bagpipes\ stellar mass bins ranging $M_{\star}=10^{8.5-10.5}~\rm M_{\odot}$ in equal width $0.5~\rm dex$ bins to investigate the origin of the scatter. Since the best-fit parameters and scatter in our star forming subsamples split by mass remains approximately consistent with the total, we focus our discussion on the scatter within our smouldering subsample, with results shown in \autoref{fig:xi_ion_MUV_mass_split}. 

Two differences can be observed for our smouldering sample. Firstly, at a given $M_{\rm UV}$ galaxies have reduced \xiionzero\ at higher $M_{\star}$, indicative of increasingly self-regulated systems with less stochastic and more secular star formation. Secondly, the $\sigma_{\xi_{\rm ion,0}}$ scatter additionally reduces with increased $M_{\star}$, seen in the lower left inset of \autoref{fig:xi_ion_MUV_mass_split}. This may favour a scenario where high mass ($M_{\star}\simeq10^{9.5-10}\rm M_{\odot}$) mini-quenched systems facilitate LyC escape only from density-bounded nebulae as a result of their reduced ISM porosities. Since more massive systems are on average older and in more overdense environments, they are likely to be gas poor, making this scenario more plausible than the alternative ``ionization bounded with holes'' which may be more common in lower mass bursty systems \citep[see e.g.][for a further discussion]{Zackrisson2013, Duncan2015}.

\section{Reionization budget constraints}
\label{sec:total_reionization_contribution}

\subsection{Formulating the contribution of galaxies to reionization}


In this work, we determine the total contribution of galaxies to reionization in terms of \MUV, which is less prone to the systematics stemming from the assumptions made when performing SED fitting than the stellar mass. While F444W NIRCam grism spectroscopy has been used to calculate the \ha\ luminosity function at $z\simeq6$ \citep{Covelo-Paz2025, Xiaojing2025} in the First Reionization Epoch Spectroscopically Complete Observations \citep[FRESCO;][]{Oesch2023} survey, the selection of these sources remains line flux limited. Using deeper JWST/NIRCam allows us to produce independent measurements in a UV selected sample, which may in theory host a significant number of smouldering galaxies missed by slitless spectroscopy.

Defining \ndotion\ as the rate of ionizing photon production into the IGM per unit co-moving volume, the total contribution from galaxies in terms of \MUV\ is thus written as
\begin{equation}
    \dot{n}_{\mathrm{ion}} = \int^{M_{\mathrm{UV, lim}}}_{-\infty}\Phi(M_{\mathrm{UV}})f_{\mathrm{esc}}(M_{\mathrm{UV}})\dot{N}_{\mathrm{ion}}(M_{\mathrm{UV}})dM_{\mathrm{UV}} ,
    \label{eq:total_ndot_ion}
\end{equation}
where $\Phi(M_{\mathrm{UV}})$ is the UVLF and $M_{\mathrm{UV, lim}}$ is the UV magnitude corresponding to the turnover of the UVLF. In the remainder of this section, we produce \MUV\ dependent scaling relations and UVLFs for our entire combined sample as well as for our star forming and smouldering subsamples, summing their contributions to determine a total.

\subsection{\texorpdfstring{$\beta_{\rm UV}-M_{\rm UV}$}{Magnitude dependence of UV slopes} and a new empirical \texorpdfstring{\fesc\ }{escape fraction }prescription}
\label{sec:EPOCHS_v2_beta_MUV}

\begin{figure}
    \centering
    \includegraphics[width=1.05\linewidth]{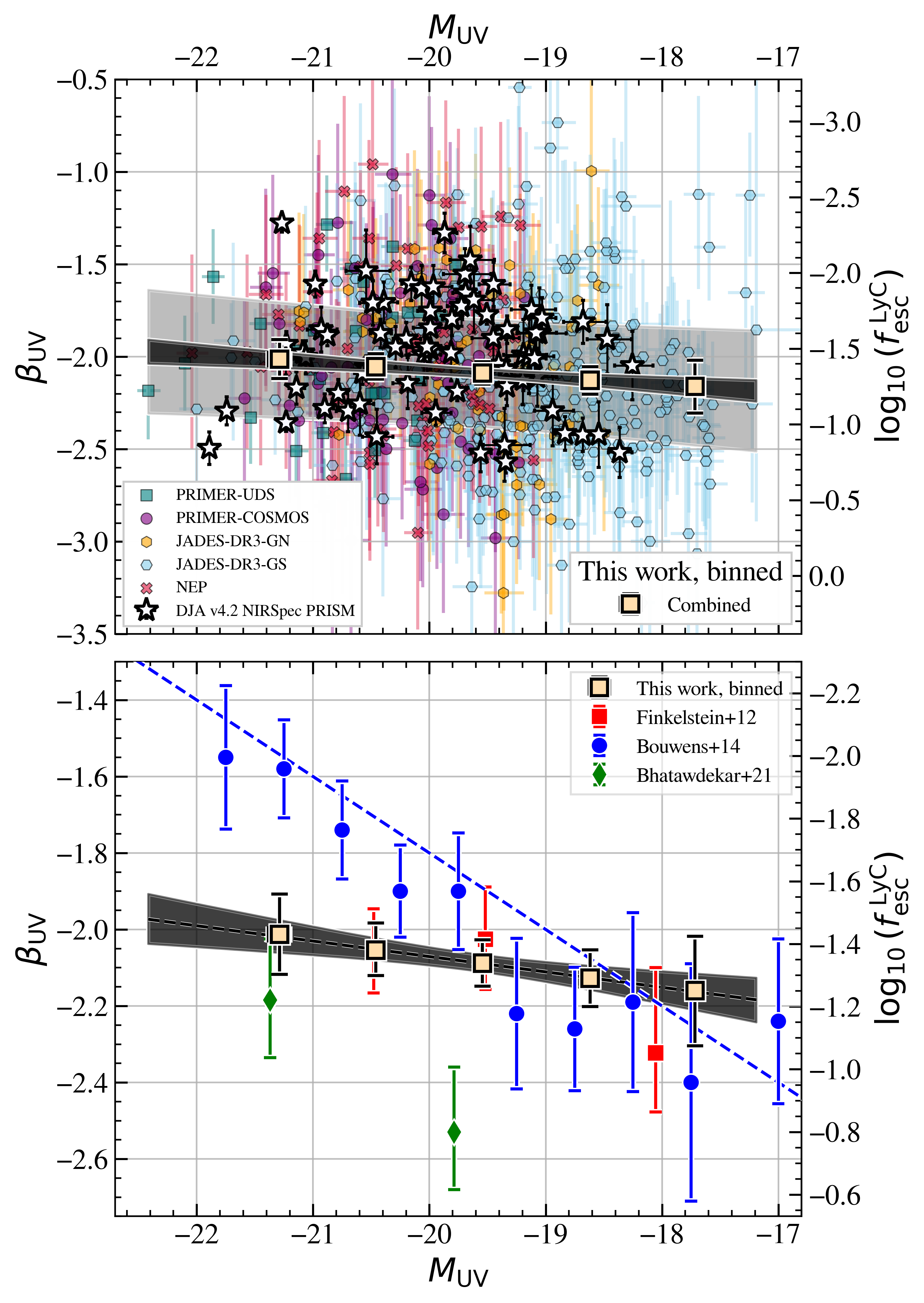}
    \caption[$\beta_{\rm UV}-M_{\rm UV}$ relation derived from our 90\% stellar mass complete sample]{\textbf{Upper panel)} $\beta_{\rm UV}-M_{\rm UV}$ relation for our 90\% stellar mass complete sample at $5.6<z<6.5$ coloured by EPOCHS-v2 origin survey. Spectroscopic sources from the DJA are shown as white stars. The dark shaded region shows the $16^{\rm th}-84^{\rm th}$ percentile of the power law fit, with the light shaded region showing the symmetric $\sigma_{\beta, \rm scat}=0.27\pm0.02$ scatter. \textbf{Lower panel)} Our best-fit model, and wild bootstrap-binned data points (beige squares), compared to \hst\ studies from \citet{Finkelstein2012}, \citet{Bouwens2014}, and \citet{Bhatawdekar2021} at $z\simeq6$. The secondary y-axis shows $\log_{10}(f_{\rm esc}^{\rm LyC})$ calculated using the LzLCS scaling relation from $\beta_{\rm UV}$ presented in \citet{Chisholm2022}.}
    \label{fig:EPOCHS_v2_beta_MUV}
\end{figure}

The dependence of $\beta_{\rm UV}$ on \MUV\ for our 90\% stellar mass complete sample is measured in \autoref{fig:EPOCHS_v2_beta_MUV}. We utilize Monte Carlo Markov Chain (MCMC) sampling techniques to fit a power law using the {\tt emcee} python package \citep{Foreman-Mackey2013}. Wide uniform priors on the amplitude and slope are adopted with limits $\pm10\sigma$ from the median value of a preliminary linear least squares fit. We model the scatter in amplitude, $\sigma_{\beta,\rm scat}$, as a constant which is additionally left free in the fitting procedure with a wide, positive, uniform prior. A mild trend of
\begin{equation}
    \beta_{\rm UV} = (-0.040\pm0.022)M_{\rm UV} -2.88\pm0.34
    \label{eq:beta_MUV}
\end{equation}
is observed such that fainter galaxies are bluer with $\sigma_{\beta, \rm scat}=0.27\pm0.02$ scatter. No significant differences in the best fit parameters and scatter are found between the star forming and smouldering subsamples.

Compared with previous HST studies utilizing ACS WFC and WFC3-IR data over the XDF, HUDF, CANDELS, and HFF lensing fields \citep[e.g.][]{Finkelstein2012, Bouwens2014, Bhatawdekar2021}, we find a somewhat shallower slope. This may be due to the fact that we use the ultra-blue ``fsps\_larson'' template set used in the \eazypy\ SED fitting procedure, however our analogous sample of spectroscopically confirmed $z\simeq6$ sources from the DJA suggest that some of these bright ultra-blue sources may be the so-called ``blue monsters'' from \citet{Ziparo2023}. Perhaps this unusual population of galaxies have large specific SFRs (sSFR) with strong radiation driven outflows which remove copious amounts of dust from the ISM \citep{Ferrara2024} or are strong leakers of LyC photons, perhaps in a different mode than leaking faint galaxies \citep{Katz2023}. 


To first order, the escape fraction of LyC from the ISM of galaxies can be determined from measurements of $\beta_{\rm UV}$, albeit with a $\gtrsim1~\rm dex$ scatter \citep{Chisholm2022}. Converting the relation in \autoref{eq:beta_MUV} to \fesc\ following this prescription, derived from the Low-redshift Lyman Continuum Survey (LzLCS), we find a fairly moderate $\langle f_{\rm esc}^{\rm LyC}\rangle\simeq5\%$ at $M_{\rm UV}\simeq-19.0$. Fainter galaxies are found to have on average slightly larger \fesc\ and vice versa, however since our $\beta_{\rm UV}-M_{\rm UV}$ is shallower than previous studies have found, the difference is somewhat reduced. We do not use multivariate models to estimate \fesc\ in this work \citep{Choustikov2024a, Choustikov2024b, Jaskot2024a, Jaskot2024b}, although this remains a promising future avenue for photometric and spectroscopic studies.

\subsection{The relationship of \texorpdfstring{$\dot{N}_{\rm ion, 0}$}{ionizing photon production rate} with \texorpdfstring{$M_{\rm UV}$}{UV magnitude}}
\label{sec:EPOCHS_v2_ndot_ion_MUV}

For each source, we compute \Ndotionzero\ directly within the \bagpipes\ ``model\_galaxy'' to produce full Bayesian posteriors following \autoref{eq:Ndot_ion} and \autoref{eq:escaping_ndot_xi_ion}. These \Ndotionzero\ values are subsequently inflated for extended sources following the same methodology as the stellar masses outlined in \autoref{sec:Halpha_bagpipes}. As before, we plot the resulting \Ndotionzero$-$\MUV\ relation in \autoref{fig:Ndot_ion_MUV}, plotting the combined photometric (split by origin survey) and spectroscopic sample in the upper panel and splitting into SFG and smouldering subsamples in the lower panel. 

The literature comparison is similar to that shown in \autoref{fig:xi_ion_MUV_lit_comp}, although we note that the \citet{Simmonds2024b} \Ndotionzero$-$\MUV\ relation is slightly flatter than our SFG sample, most likely due to their dust assumptions made when SED fitting with \prospector. Our data has a larger dynamic range in UV magnitude, especially at the bright end, due to our inclusion of data from PRIMER-UDS, PRIMER-COSMOS, and NEP fields, as opposed to the JADES (GOODS-South + GOODS-North) data used in \citet{Simmonds2024b}.

Although clearly visible in the \xiionzero$-$\MUV\ relation in \autoref{fig:EPOCHS_v2_xi_ion_MUV}, the upper envelope of the prior in \xiionzero\ is not as apparent in \Ndotionzero$-$\MUV\ space. The likely culprit is the dust attenuation; the UV luminosity on the denominator of \autoref{eq:xi_ion_definition} is dust-corrected, which means the upper \Ndotionzero$-$\MUV\ prior envelope no longer follows a power-law since \MUV\ is an observed (non dust-corrected) quantity.

We fit power-law relations to the data in \autoref{fig:Ndot_ion_MUV} in a similar way to the \xiionzero$-$\MUV\ data, additionally including an \MUV\ dependence in the scatter to account for the heteroscedasticity which now exists in this variable (i.e. sources that are faint in $L_{\nu, \rm UV}$ are likely also faint in $L_{\rm H\alpha}$). Results of this fitting procedure are presented in \autoref{tab:power_law_params} in \autoref{sec:power_law_params}. Wild bootstrap-binned data are computed following \autoref{eq:wild_bootstrapping} replacing $\xi_{\mathrm{ion,0}, i}$ with $\dot{N}_{\mathrm{ion,0}, i}$.

\begin{figure}
    \centering
    \includegraphics[width=0.98\linewidth]{figures/Ndot_ion_MUV.png}
    \caption{$\dot{N}_{\rm ion}-M_{\rm UV}$ relation for our 90\% stellar mass complete $5.6<z<6.5$ sample in the large area EPOCHS-v2 surveys. The dark shaded regions show the $16^{\rm th}-84^{\rm th}$ percentile of the power law fit to each subsample, with the light shaded regions showing the asymmetric scatter fitted as a function of both $\dot{N}_{\rm ion}$ and \MUV. \textbf{Upper panel)} Data is coloured by EPOCHS-v2 origin survey. Spectroscopic sources from the DJA are shown as white stars and wild bootstrap-binned data points are shown as beige squares. \textbf{Lower panel)} Results for our star forming (blue) and smouldering (red) subsamples. Contours are shown for $\{1,2,3\}\sigma$ significance levels. Wild-bootstrap-binned data is shown as light blue/light red circles.}
    \label{fig:Ndot_ion_MUV}
\end{figure}

\subsection{UV luminosity function}

\subsubsection{Calculating the UVLF using the \texorpdfstring{$1/V_{\rm max}$}{1/Vmax} method}
\label{sec:calculating_UVLF}

Taking our final sample of $1721$ galaxies, we construct the UVLF using the $1/V_{\rm max}$ method \citep{Schmidt1968, rowanrobinson1968} at $z\simeq6$ for the first time with a JWST/NIRCam selected sample. For each galaxy, we calculate the maximum ($z_{\rm max}$) and minimum ($z_{\rm min}$) detectable redshift in each field using the average depth measurements combined with the selection criteria listed in \autoref{sec:EPOCHS_v2_sample_selection} in steps of $\delta z = 0.01$. From these $z_{\rm min}$ and $z_{\rm max}$ values, we calculate the comoving $V_{{\rm max}, i, j}$ for each galaxy, $i$, in each field, $j$, in our assumed cosmology using the unmasked areas given in \autoref{sec:data_and_sample}. The maximum detectable volume for each galaxy ($V_{{\rm max}, i}$) can be determined by simply summing the $V_{{\rm max}, i, j}$ for each field.

\begin{figure*}
    \centering
    \includegraphics[width=0.75\linewidth]{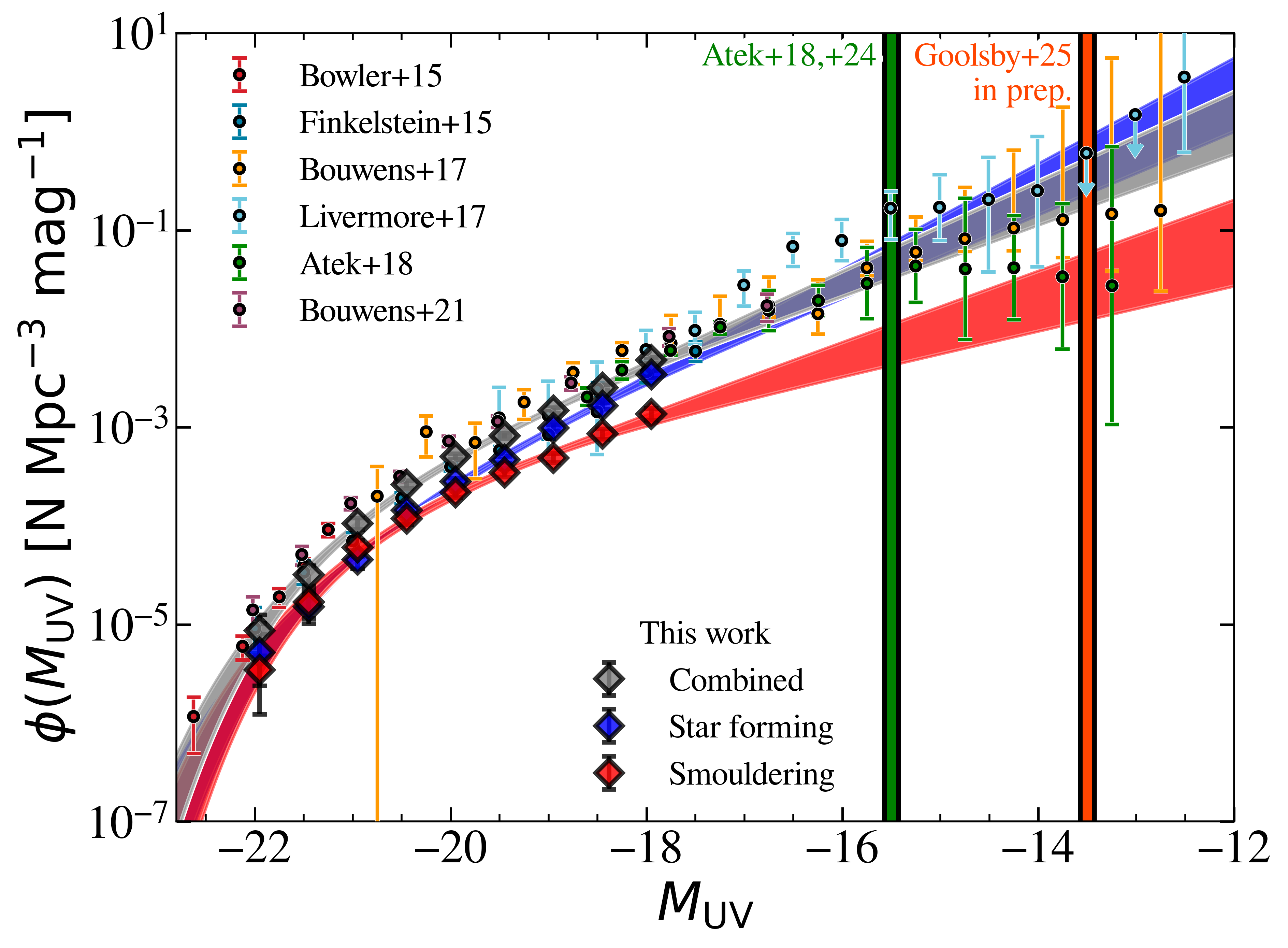}
    \caption[$z\simeq6$ UVLF for the combined sample as well as our star forming and smouldering subsamples]{$z\simeq6$ UVLF for the combined, star forming, and smouldering subsamples plotted as diamonds. \citet{Schechter1976} model posterior solutions are plotted for each subsample, with the shaded region highlighting the $1\sigma$ errors. Results are compared to measurements of the bright end \citep[][from the ground-based UltraVISTA survey]{Bowler2015}, the knee \citep[][from \hst\ blank fields]{Finkelstein2015, Bouwens2021}, and the faint end\citep[][from the HFF strong lensing clusters]{Bouwens2017, Livermore2017, Atek2018}. Approximate turnover magnitudes from \jwst-derived measurements by \citet{Atek2024} and Goolsby et al. (in prep.) are shown as thick vertical green and orange lines respectively.}
    \label{fig:uvlf_fit}
\end{figure*}

The sample is split into magnitude bins of size $\Delta M_{\rm UV}=0.5$ and the number density in each bin is calculated as
\begin{equation}
    \Phi(M_{\rm UV})d\log_{10}(M_{\rm UV}) = \frac{1}{\Delta M_{\rm UV}}\sum_{i}\frac{1}{C_{i}V_{{\rm max}, i}},
    \label{eq:UVLF}
\end{equation}
where $C_{i}$ is the completeness of galaxy $i$ calculated as a function of F115W SNR (corresponding to the rest-frame UV at $z\simeq6$) following the procedure explained in \autoref{sec:EPOCHS_v2_sample_completeness}. The Poisson uncertainty in each bin is calculated as
\begin{equation}
    \delta\Phi(M_{\rm UV}) = \frac{1}{\Delta M_{\rm UV}}\sqrt{\sum^N_{i}\Bigg(\frac{1}{C_{i}V_{{\rm max}, i}}\Bigg)^2}.
\end{equation}
For magnitude bins with low occupancy ($N\leq4$) the above formula overestimates the error. In these instances, we adopt the Poisson confidence interval from \citet{Ulm1990} based on the $\chi^2$ distribution $I=[0.5\chi^2_{2N, a/2}, 0.5\chi^2_{2(N+1), 1-a/2}]$ similarly to \citet{Adams2024}, \citet{Adams2025}, and \citet{Harvey2025}. The UVLF for our sample, as well as our star forming and smouldering subsamples, is shown in \autoref{fig:uvlf_fit} which includes completeness corrections and cosmic variance as explained below.

\begin{table}[]
   \centering
   \caption{Completeness corrected UV luminosity function bins for the combined, star forming, smouldering samples, as plotted in \autoref{fig:uvlf_fit}. Amplitudes are measured as per \autoref{eq:UVLF} and errors comprise of both Poisson and cosmic variance derived via the prescription given in \citet{Driver2010} assuming square survey geometries in each case.}
   \setlength{\tabcolsep}{2pt}
   \begin{tabular}{ccc}
   \toprule
    $M_{\rm UV}$ bin & $\Phi(M_{\rm UV})$ & \multirow{2}{*}{$N_{\rm gals}$} \\
    (AB mag) & ($\times10^{-4}~\rm mag^{-1}cMpc^{-3}$) & \\
   \hline
   \multicolumn{3}{c}{Combined sample} \\
    $[-22.25,-21.75]$ & $0.08\pm0.04$ & $5$ \\
   $[-21.75,-21.25]$ & $0.32\pm0.07$ & $19$ \\
   $[-21.25,-20.75]$ & $1.06\pm0.14$ & $60$ \\
   $[-20.75, -20.25]$ & $2.63\pm0.22$ & $149$ \\
   $[-20.25, -19.75]$ & $5.03\pm0.34$ & $263$ \\
   $[-19.75, -19.25]$ & $8.19\pm0.50$ & $318$ \\
   $[-19.25, -18.75]$ & $14.84\pm0.91$ & $312$ \\
   $[-18.75, -18.25]$ & $25.21\pm1.72$ & $280$ \\
   $[-18.25, -17.75]$ & $48.07\pm3.41$ & $213$ \\
   \hline
   \multicolumn{3}{c}{Star forming} \\
    $[-22.25,-21.75]$ & $0.05_{-0.03}^{+0.05}$ & $3$ \\
   $[-21.75,-21.25]$ & $0.15\pm0.05$ & $9$ \\
   $[-21.25,-20.75]$ & $0.45\pm0.09$ & $27$ \\
   $[-20.75, -20.25]$ & $1.45\pm0.16$ & $83$ \\
   $[-20.25, -19.75]$ & $2.83\pm0.23$ & $156$ \\
   $[-19.75, -19.25]$ & $4.69\pm0.37$ & $187$ \\
   $[-19.25, -18.75]$ & $9.85\pm0.75$ & $207$ \\
   $[-18.75, -18.25]$ & $16.58\pm1.37$ & $188$ \\
   $[-18.25, -17.75]$ & $34.41\pm2.87$ & $153$ \\
   \hline
   \multicolumn{3}{c}{Smouldering} \\
    $[-22.25,-21.75]$ & $0.03_{-0.02}^{+0.05}$ & $2$ \\
   $[-21.75,-21.25]$ & $0.17\pm0.05$ & $10$ \\
   $[-21.25,-20.75]$ & $0.60\pm0.11$ & $33$ \\
   $[-20.75, -20.25]$ & $1.18\pm0.15$ & $66$ \\
   $[-20.25, -19.75]$ & $2.18\pm0.24$ & $107$ \\
   $[-19.75, -19.25]$ & $3.47\pm0.34$ & $131$ \\
   $[-19.25, -18.75]$ & $4.89\pm0.52$ & $105$ \\
   $[-18.75, -18.25]$ & $8.58\pm1.04$ & $92$ \\
   $[-18.25, -17.75]$ & $13.66\pm1.84$ & $60$ \\
   \botrule
   \end{tabular}
   \label{tab:uvlf_measurements}
\end{table}

\subsubsection{Estimating completeness}
\label{sec:UVLF_completeness}


The completeness of the various fields used in this sample is shown in \autoref{sec:EPOCHS_v2_sample_completeness}, where for the UVLF, we use the completeness curves in terms of \MUV\ rather than stellar mass. Since our surveys are rarely completely uniform in depth, we define our completeness in bins of F115W SNR corresponding to the rest-frame UV at $z\simeq6$; correction factors, $C_{i,j}$, are calculated for each galaxy in each survey. We find that our faintest \MUV\ bin is $\lesssim30-50\%$ complete, which we decide not to use in the Schechter fitting procedure explained below.



The detection completeness of smouldering galaxies is expected to better scale with peak $\Sigma_{\rm SFR_{10 Myr}}$ \citep{Endsley2024}, meaning it is quite likely that these galaxies suffer more from detection incompleteness than our star forming subsample. A thorough analysis using high resolution simulations such as FIRE-2 \citep[][]{Hopkins2018} or SPHINX \citep[][]{Katz2023_sphinx} is required in order to determine and correct for the full extent of this incompleteness to better constrain the faint end slope of the smouldering galaxy UVLF.


\subsubsection{Cosmic variance}
\label{sec:cosmic_variance}

In addition to the Poisson errors, we additionally include errors from cosmic variance using the prescription in \citet{Driver2010} and combine the cosmic variance measurements made in each survey using the weighted sum of squares from \citet{Moster2011}. We find that our total cosmic variance measurement at $z\simeq6$ is $11\%$. It is worth noting that this value is approximate at this redshift and relies on measurements of galaxy biases from clustering across wide area surveys, which are expected to be strongly dependent on stellar mass and UV luminosity \citep[e.g.][]{Zehavi2011, Harikane2021}. Upcoming galaxy clustering measurements from Euclid should greatly improve the accuracy and precision of these estimates.


\subsubsection{Fitting results}
\label{sec:UVLF_fitting_results}

\begin{figure}
    \centering
        \begin{subfigure}[b]{0.48\textwidth}
        \centering
            \includegraphics[width=\linewidth]{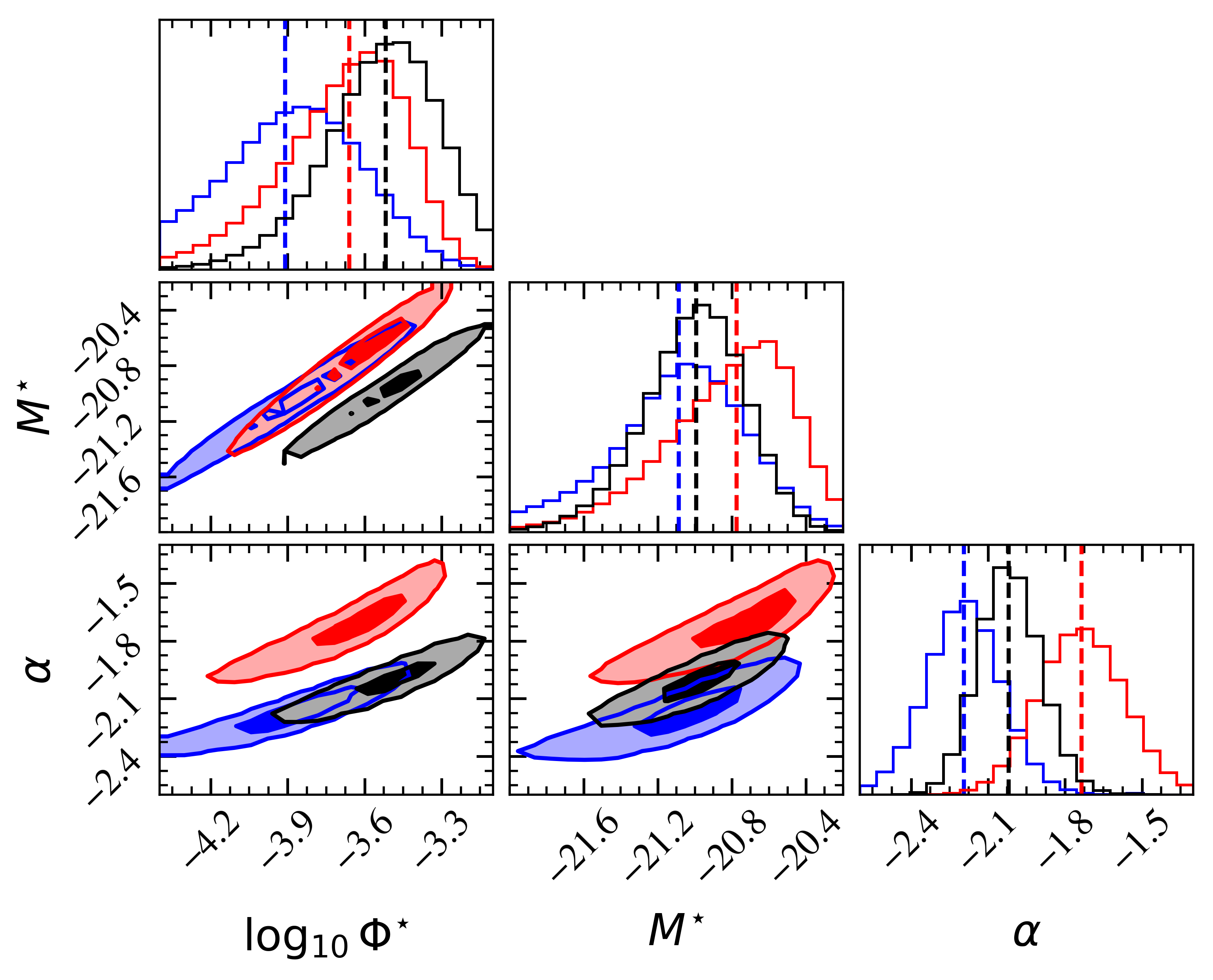}
        \end{subfigure}
        \hfill
        \hfill
        \begin{subfigure}[b]{0.48\textwidth}
        \centering
            \includegraphics[width=\linewidth]{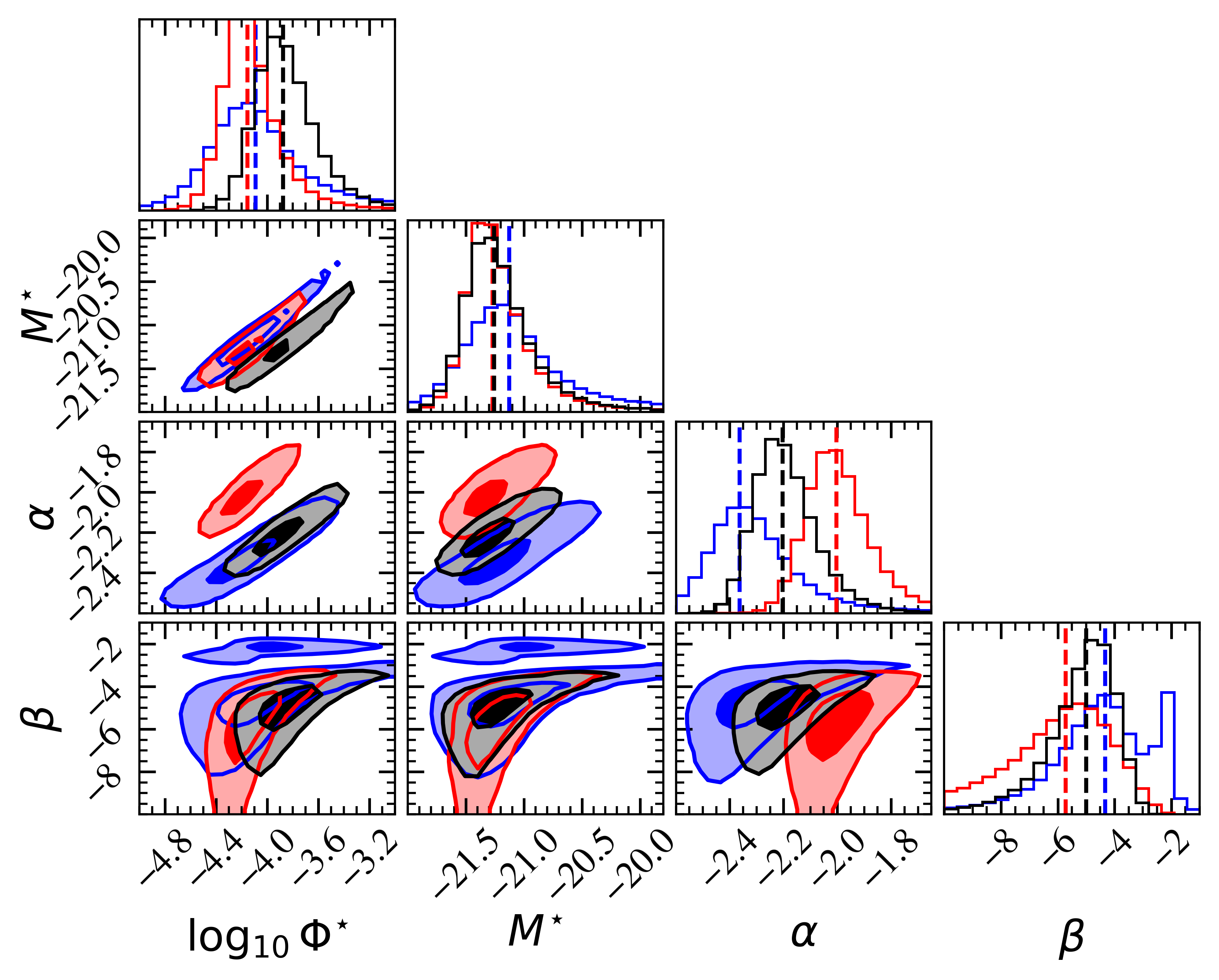}
        \end{subfigure}
    \caption[\citet{Schechter1976} and DPL posterior corner plots for our combined, star forming, and smouldering UVLFs at $z\simeq6$]{Posterior corner plots for fits to our combined (black), star forming (blue), and smouldering (red) UVLFs at $z\simeq6$. Dashed vertical lines show the median of each parameter distribution for each subsample. \textbf{Upper panel)} \citet{Schechter1976} model with $\Theta=\{\log_{10}\Phi^{\star}, M^{\star}, \alpha\}$ free parameters. \textbf{Lower panel)} DPL model with $\Theta=\{\log_{10}\Phi^{\star}, M^{\star}, \alpha, \beta\}$ free parameters}
    \label{fig:uvlf_corner}
\end{figure}

We fit our measured $z\simeq6$ UVLFs with a \citet{Schechter1976} luminosity function utilizing a Bayesian MCMC fitting procedure using the {\tt emcee} python package \citep{Foreman-Mackey2013}, displaying the resulting corner plot in the upper panel of \autoref{fig:uvlf_corner}. For consistency, we first check whether fitting results for the sample as a whole agree with previous Schechter luminosity function estimates. Our results are compared to the ground-based UltraVISTA \citep{McCracken2012} study by \citet{Bowler2015}, blank field \hst\ imaging results \citep{Finkelstein2015, Bouwens2021}, and the strongly lensed Hubble Frontier Field \citep[HFF;][]{Bouwens2017, Livermore2017, Atek2018}.

We find that the amplitude, $\Phi^{\star}$, is slightly lower than the $\log_{10}\Phi^{\star}=-3.1$ measured in \citet{Bouwens2021}, however it remains consistent with a similar study by \citet{Finkelstein2015}, $\log_{10}\Phi^{\star}=-3.7$ across the same field areas. The measured faint-end slopes of $\alpha=-1.93\pm0.08$ \citep{Bouwens2021} and $\alpha=-2.02\pm0.10$ \citep{Finkelstein2015} at $z\simeq6$ in the literature are consistent with our fitting results of $\alpha=-2.0\pm0.2$. As expected, the amplitudes of our subsample UVLF fits are slightly lower than the combined UVLF since there are fewer galaxies in our star forming and smouldering subsamples compared to the combined sample, with $M^{\star}$ remaining approximately consistent. The largest difference between our subsamples lies in the measured faint-end slope, for which we obtain $\alpha_{\rm SFG}=-2.2\pm0.2$ and $\alpha_{\rm sm}=-1.7\pm0.2$. 

Since a double power law (DPL) fit has been shown to better match the smooth evolution at the bright end of the UVLF \citep{Bowler2014, Bowler2015, Bowler2017, Bowler2020}, we additionally fit a DPL to our total, star forming, and smouldering UVLFs. Wide uniform priors on the $\Theta=\{\Phi^{\star}, M^{\star}, \alpha, \beta\}$ free parameters are used in the {\tt emcee} fitting procedure, where $\beta$ is the bright end slope found to be $\beta=-4.4$ by \citet{Bowler2014} at $z=6$. The posterior corner plot is shown in the lower panel of \autoref{fig:uvlf_corner}, where wide $\beta$ posteriors for each subsample (as well as the total) and a bimodal solution for $\beta_{\rm SFG}$ is observed, highlighting the need for larger area surveys to properly sample this regime. Nevertheless, our results remain consistent at the faint end, which appears robust to our model choice. 

To quantify which model is more appropriate to use, we calculate both the Akaike Information Criterion (AIC) and Bayesian Information Criterion (BIC) for both \citet{Schechter1976} and DPL UVLF fits. For the total sample, we find $\Delta\rm AIC=\mathrm{AIC}_{DPL} - \mathrm{AIC}_{Sch}=0.82$ and $\Delta\rm BIC=\mathrm{BIC}_{DPL} - \mathrm{BIC}_{Sch}=1.02$, highlighting a slight, but insignificant, favour towards the \citet{Schechter1976} model due to its simplicity. The same conclusion is reached for our star forming ($\Delta\rm AIC=1.35$; $\Delta\rm BIC=1.55$) and smouldering ($\Delta\rm AIC=0.51$; $\Delta\rm BIC=0.71$) subsamples.

\subsection{Reionization contribution as a function of \texorpdfstring{\MUV}{UV magnitude}: Tying it all together}
\label{sec:tying_it_all_together}

\begin{figure}
    \centering
    \includegraphics[width=0.95\linewidth]{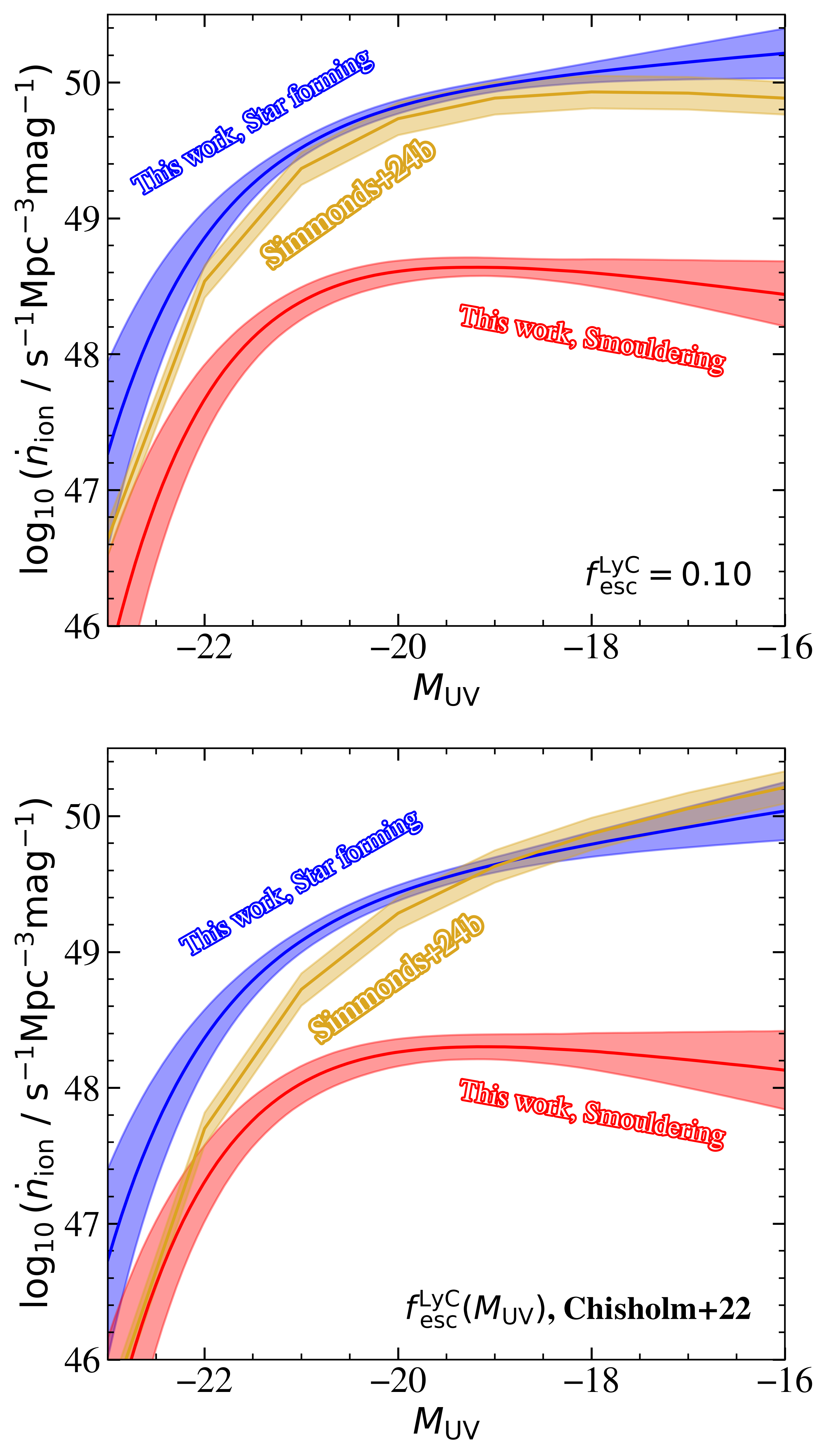}
    \caption[Total contribution to reionization as a function of \MUV\ for our combined, star forming, and smouldering subsamples]{Total contribution to reionization as a function of \MUV\ for our star forming (blue), and smouldering (red) subsamples compared to the \citealt{Simmonds2024b} results (yellow) using the same \fesc\ parametrization. The total contribution closely resembles the star forming subsample, which dominates the ionizing photon budget, and is therefore not shown. \textbf{Upper panel)} Results use fixed $f_{\rm esc}^{\rm LyC}=10\%$. \textbf{Lower panel)} Results use \citet{Chisholm2022} derived $f_{\rm esc}^{\rm LyC}(M_{\rm UV})$.}
    \label{fig:ndot_ion_integrand}
\end{figure}

The total rate of production of ionizing photons escaping into the IGM is calculated following \autoref{eq:total_ndot_ion}, the integrand of which is shown as a function of \MUV\ in \autoref{fig:ndot_ion_integrand} both for \citet{Chisholm2022} derived $f_{\rm esc}^{\rm LyC}(M_{\rm UV})$ and fixed $f_{\rm esc}^{\rm LyC}=10\%$. In both cases, the contribution from SFGs is $>1~\rm dex$ larger than smouldering galaxies, which reduces as with fainter \MUV, potentially due to reduced selectability of smouldering galaxies towards the detection limit of our photometric datasets. The SFG contours additionally highlight that faint galaxies that are currently in a starburst phase contribute the majority of the ionizing photons during the EoR.

For fixed $f_{\rm esc}^{\rm LyC}=10\%$, our SFG subsample contributes $\sim0.2-0.5~\rm dex$ more than the \citet{Simmonds2024b} sample, while retaining a similar shape. The amplitude is thus a function of UVLF and \Ndotionzero\ normalizations. Our power law fits to the \Ndotionzero$-$\MUV\ relation for SFGs displays a higher amplitude than the JADES (GOODS-North + GOODS-South) results from \citet{Simmonds2024b}. They adopt the \citet{Bouwens2021} UVLF, which has a larger normalization and shallower faint-end slope than our SFG UVLF shown in \autoref{fig:uvlf_fit}. This opposes our larger \Ndotionzero\ measurements and leads to our sample producing an increasing amount faintwards of the UVLF ``knee'' compared to \citet{Simmonds2024b}. 


In this work, we measure a flatter $\beta_{\rm UV}-M_{\rm UV}$ than the work of \citet{Simmonds2024b}. The impact on \ndotion\ can be seen when adopting the \citet{Chisholm2022} \fesc\ parametrization in the lower panel of \autoref{fig:ndot_ion_integrand}. This increases the prominence of bright galaxies to reionization. Since \fesc\ models do not commonly treat SFGs and smouldering galaxies distinctly, our treatment does not highlight the potential differences between LyC escape pathways for different subsamples of galaxies. We note that results assuming alternative \fesc\ parametrizations, not shown in this work, may increase the contribution from the brightest ``oligarchs'' \citep[e.g.][]{Naidu2020} and/or change the amplitude of the curves in \autoref{fig:ndot_ion_integrand}. 

\subsection{Reionization constraints at \texorpdfstring{$z\simeq6$}{z=6}}

To determine an estimate of the extent to which UV selected galaxies contribute to reionization, we rather arbitrarily define the UVLF turnover to be at $M_{\rm UV, lim}=-17$. Given this rather crude assumption, we find that our sample as a whole generates $\log_{10}\dot{n}_{\rm ion}=50.68\pm 0.03$ and $\log_{10}\dot{n}_{\rm ion}=50.34\pm 0.04$ for \fesc$=0.1$ and \citet{Chisholm2022} \fesc\ respectively, and that in both cases the smouldering subsample accounts for $<10\%$ of the combined budget.

\begin{figure*}
    \centering
    \includegraphics[width=0.98\linewidth]{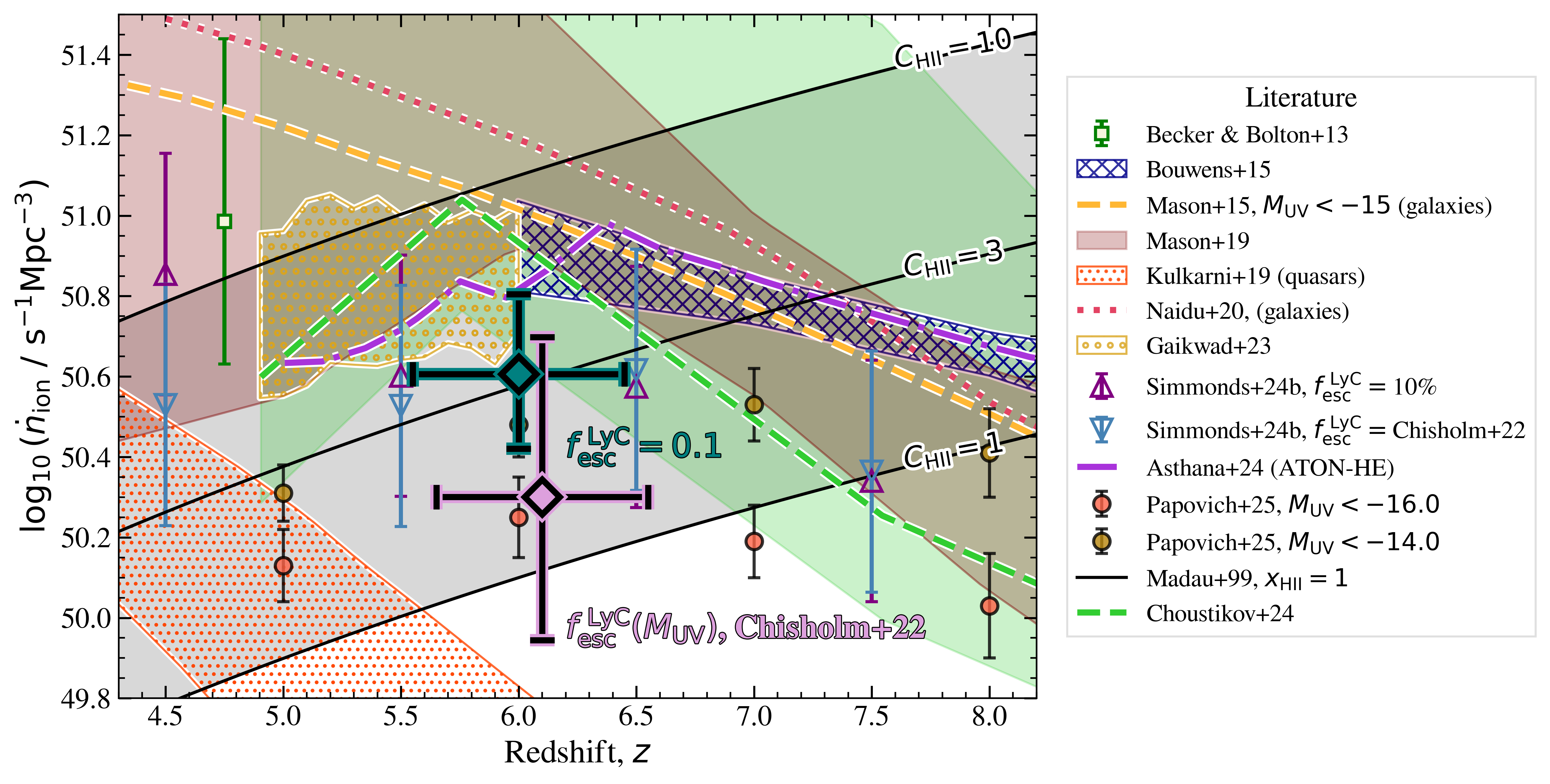}
    \caption[Total rate of ionizing photon production integrated to $M_{\rm UV, lim}=-17.0$ for our two \fesc\ prescriptions]{Total rate of ionizing photon production integrated to $M_{\rm UV, lim}=-17.0$. Results are compared to observations of galaxies \citep{Becker2013, Bouwens2015-EoR, Mason2015b, Mason2019, Naidu2020, Gaikwad2023, Simmonds2024b, Choustikov2025, Papovich2025}, AGN \citep{Kulkarni2019}, and the ATON-HE simulation \citep{Asthana2024}. Results adopting our \MUV\ dependent escape fractions derived from the \citet{Chisholm2022} scaling relation from $\beta_{\rm UV}$ are shown as a pink diamond, shifted by $\Delta z=+0.1$ for clarity, and our $f_{\rm esc}^{\rm LyC}=10\%$ result is in teal.}
    \label{fig:Ndot_ion_total}
\end{figure*}

Results of this integration are shown in \autoref{fig:Ndot_ion_total}, where we additionally compare to observational constraints from galaxies \citep{Becker2013, Bouwens2015, Mason2015b, Mason2019, Naidu2020, Gaikwad2023, Simmonds2024b, Choustikov2025, Papovich2025}, as well as inferences from the AGN luminosity function \citep{Kulkarni2019}, and the late reionization model from the ATON-HE reionization simulation \citep{Asthana2024}. Contours showing the required ionizing budget to produce a fully stably ionized Universe with given volume-averaged \ion{H}{2} clumping factors, \cHII\,$=\{1,3,10\}$, are calculated following the prescription in \citet{Madau1999},
\begin{equation}
    \dot{n}_{\rm ion} = 10^{51.2}\bigg(\frac{C_{\rm HII, rec}}{30}\bigg)\bigg(\frac{1+z}{6}\bigg)^3\bigg(\frac{\Omega_{\rm b}h_{50}^2}{0.08}\bigg)^2 ~[\rm s^{-1}Mpc^{-3}],
\end{equation}
and are additionally plotted in \autoref{fig:Ndot_ion_total}.

Consistent results are found with the observational studies in the literature for fixed $f_{\rm esc}^{\rm LyC}=10\%$, however we measure an $\dot{n}_{\rm ion}\sim0.3~\rm dex$ lower adopting the \citet{Chisholm2022} \fesc\ relation from $\beta_{\rm UV}$ due to our flatter $\beta_{\rm UV}-M_{\rm UV}$ relation which produces on average $f_{\rm esc}^{\rm LyC}\simeq5\%$. 

\section{Implications for IGM \ion{H}{2} clumping}
\label{sec:clumping_factor_implications}

\subsection{Hydrogen reionization formulation}
\label{sec:hydrogen_reionization_formulation}

The reionization of hydrogen has long since been modelled analytically in terms of the ionized fraction, $x_{\rm HII}\equiv n_{\rm HII}/n_{\rm H}$, defined in terms of the number densities of both atomic (HI), and ionized (HII) hydrogen such that $n_{\rm H}=n_{\rm HI} + n_{\rm HII}$. The evolution of this $x_{\rm HII}$ is commonly modelled via the volume-filling factor of HII, $Q_{\rm HII}$. This is approximately equivalent to $x_{\rm HII}$ pre-overlap of ionized bubbles; $Q_{\rm HII}>x_{\rm HII}$ post-overlap due to residual neutral hydrogen remaining within ionized regions. The evolution of $Q_{\rm HII}$ is modelled following the simple ordinary differential equation (ODE) prescription of \citet{Madau1999},
\begin{equation}
    \frac{\mathrm{d}Q_{\rm HII}}{\mathrm{d}t} = \frac{\dot{n}_{\rm ion}}{\langle n_{\rm H}\rangle} - \frac{Q_{\rm HII}}{\langle t_{\rm rec}\rangle},
    \label{eq:reionization_ode}
\end{equation}
where $\langle t_{\rm rec}\rangle$ is the volume-averaged recombination timescale (denoted by triangular brackets) given by
\begin{equation}
    \langle t_{\rm rec}\rangle=\Big[\langle1 + x_{\rm He}\rangle\langle n_{\rm H}\rangle\langle\alpha_{\rm rec}(T_e)\rangle C_{\rm HII, rec}\Big]^{-1},
    \label{eq:trec}
\end{equation}
in terms of the electron temperature dependent recombination co-efficient, $\alpha_{\rm rec}(T_e)$, and recombination-weighted \textit{global} clumping factor, $C_{\rm HII, rec}$. This equation accounts for the contribution of electrons available for recombination with \ion{H}{2} from singly-ionized helium, HeII, with ionization energy $24.6~\rm eV$. At $z\simeq6$ we are in the regime where all helium is singly ionized before the epoch of $\rm HeII$ reionization, which occurs at $z\sim3-4$ \citep{Wyithe2002,Gnedin2022}, meaning this additional electron density amounts to $x_{\rm He}\equiv n_{\rm He}/n_{\rm H}=Y/[4(1-Y)]\simeq0.08$.

For hydrogen reionization equilibrium, where the total rates of ionization and recombination equate (i.e. $\mathrm{d}Q_{\rm HII}/\mathrm{d}t=0$), the fraction of the Universe that can remain stably ionized is thus given by
\begin{equation}
    Q_{\rm HII} = \frac{\dot{n}_{\rm ion}(1+z)^3}{\langle n_{\rm H}\rangle^2\langle\alpha_{\rm rec}(T_e)\rangle C_{\rm HII, rec}}\frac{4(1-Y)}{(4-3Y)}.
    \label{eq:stable_reionization}
\end{equation}
In this work, we assume case B recombination and an electron temperature $T_e=10^4$~K typical of \ion{H}{2} regions in star-forming galaxies\footnote{At these temperatures the electron density, $n_e$, makes a minimal impact on $\alpha_{\rm B}$}, leading to a recombination co-efficient $\alpha_{\rm rec}\equiv\alpha_{\rm B}=2.59\times10^{-13}~\rm cm^3s^{-1}$ \citep{Osterbrock2006}. In the remainder of this paper, we use \autoref{eq:stable_reionization} to infer the \cHII\ IGM clumping factor, and discuss this in the context of cosmic reionization.


\subsection{Indirect clumping factor constraints}



\begin{figure}
    \centering
    \includegraphics[width=1.02\linewidth]{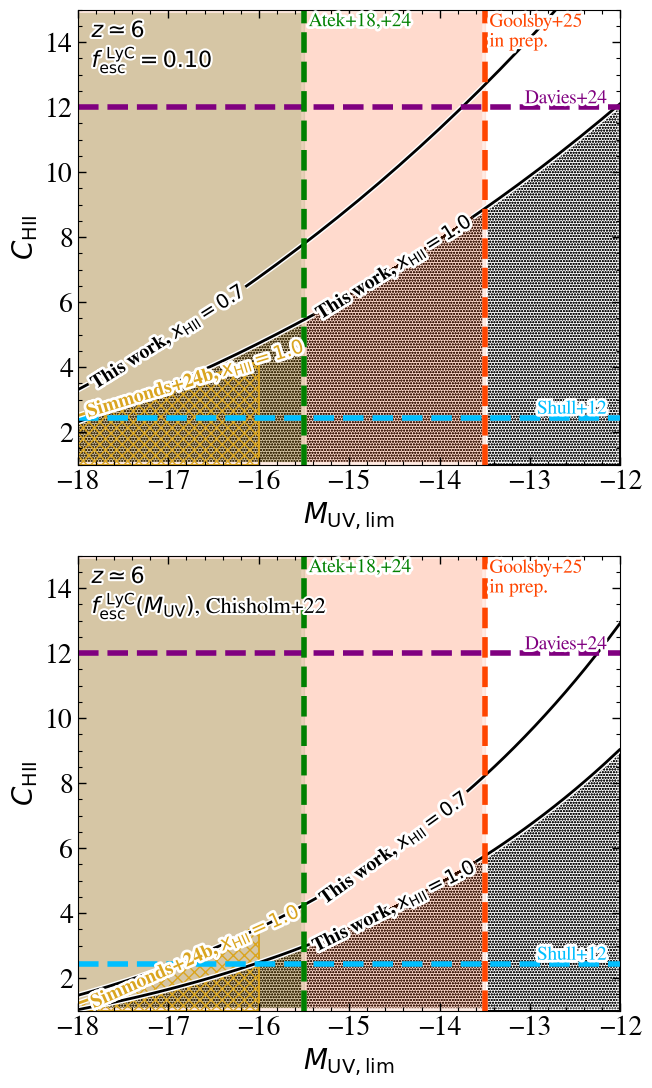}
    \caption[IGM \ion{H}{2} clumping factor constraints as a function of UVLF turnover magnitude, $M_{\rm UV, lim}$, at $z\simeq6$ for $f_{\rm esc}^{\rm LyC}(M_{\rm UV})$ and fixed $f_{\rm esc}^{\rm LyC}=10\%$]{IGM HII clumping factor constraints as a function of UVLF turnover magnitude, $M_{\rm UV, lim}$, at $z\simeq6$. Clumping factor constraints from simulations by \citet{Shull2012} (light blue) and \citet{Davies2024} (purple), and approximate observational $M_{\rm UV,lim}$ limits at $z\sim6-7$ by \citealt{Atek2018, Atek2024} (green) and  are shown. $x_{\rm HII}$ contours showing the contribution from the full (combined) sample as well as from SFGs in our work are shown in black and blue respectively, while the full sample results from \citet{Simmonds2024b} is shown in yellow. \textbf{Upper panel)}  Contours assume a fixed \fesc$=10\%$. \textbf{Lower panel)} Contours assume an \MUV\ dependent \fesc\ derived from the \citet{Chisholm2022} LzLCS conversion from $\beta_{\rm UV}$.}
    \label{fig:cHII_MUVlim}
\end{figure}

We propagate the $\dot{n}_{\rm ion}(M_{\rm UV})$ curves from \autoref{fig:ndot_ion_integrand}, extrapolated to \MUV\,$=-12$, through \autoref{eq:stable_reionization} to determine the total fraction of the Universe's hydrogen that is stably ionized as a function of $C_{\rm HII, rec}$ and UVLF turnover, $M_{\rm UV, lim}$, shown in \autoref{fig:cHII_MUVlim}. Curves for \ion{H}{2} fractions, $x_{\rm HII}\simeq Q_{\rm HII}=\{0.7, 1.0\}$ are shown, and although the neutral content of the Universe at $z\simeq6$ is dependent on whether reionization occurs early \citep[e.g.]{Becker2001, Fan2006, Yang2020} or late \citep[e.g.][]{Keating2020, Bosman2022, Keating2024, Zhu2024}, $Q_{\rm HII}$ is expected to fall within these bounds. While we obtain similar results to \citet{Simmonds2024b} at $M_{\rm UV, lim}\lesssim-17$, the largest differences appear when extrapolating to fainter \MUV.




Simulations suggest the UVLF flattens and turns over due to suppression of faint galaxy formation by reionization. The limit at which this occurs, however, is not well known. The $\mathrm{H}_2$ model of \citet{Jaacks2013} suggests this could occur at early as $M_{\rm UV, lim}=-16.0$, although the {\tt DRAGONS} \citep{Liu2016} simulation series predict a flattening at $-14$ and full turnover at $-12\lesssim M_{\rm UV}\lesssim-10$ \citep{Angel2016, Mutch2016, Poole2016}. The CoDa II simulations also suggest that this turnover occurs at $-12\lesssim M_{\rm UV}\lesssim-11$ \citep{Ocvirk2020, Dawoodbhoy2023}. 

The most recent strong-lensing results from the GLIMPSE survey at $z\simeq6$ by Goolsby et al. (in prep.) suggest no turnover as faint as $M_{\rm UV, lim}=-13.5$. Adopting this UVLF integration limit, we require $C_{\rm HII, rec}=8.8^{+4.3}_{-2.5}$ ($f_{\rm esc}^{\rm LyC}=10\%$) or $C_{\rm HII, rec}=6.2^{+4.1}_{-2.1}$ (\citet{Chisholm2022} \fesc) for a fully stably ionized Universe (i.e. with $Q_{\rm HII}=1$ and $\dot{Q}_{\rm HII}=0$), corresponding to a departure from the \citet{Shull2012} model at $2.5\sigma$ and $1.8\sigma$ significance respectively. The Bayes factor, $B=\mathcal{L}_{\rm Shull+12}/\mathcal{L}_{\rm Davies+24}$, denoting the favourability of the \citet{Davies2024} \cHII$=12$ value over the \citet{Shull2012} model assuming a fixed $M_{\rm UV, lim}=-13.5$, is measured in each case to be $\ln{B}=-5.36$ and $\ln{B}=-0.52$. The UV turnover required to favour the \citet{Davies2024} model (i.e. the limit when $\ln{B}=0$) are $M_{\rm UV, lim}>-14.2$ and $M_{\rm UV, lim}>-12.7$. 

If we were instead to assume a late-reionization model with $Q_{\rm HII}(z=6)=0.7$ and retain $M_{\rm UV, lim}=-13.5$, which remains fully consistent with direct $Q_{\rm HII}$ constraints \citep[see e.g.][for dark pixel fraction constraints]{McGreer2015,Jin2023,Davies2025}, we would need $C_{\rm HII, rec}=12.6^{+6.1}_{-3.5}$ ($f_{\rm esc}^{\rm LyC}=0.1$) or $C_{\rm HII, rec}=8.9^{+5.8}_{-3.1}$ (\citet{Chisholm2022} \fesc), providing even more significant $2.9\sigma$ and $2.1\sigma$ deviations. This reduces the turnover limit required to favour the \citet{Davies2024} results for $M_{\rm UV, lim}>-15.7$ and $M_{\rm UV, lim}>-13.7$. We thus conclude a moderate favourability of the \citet{Davies2024} model except in the case of low \fesc$\lesssim5\%$.



\begin{table*}[t]
    \centering
    \makebox[\textwidth]{}
    \begin{tabular}{c|c|cc|ccc}
    \toprule 
    \multirow{2}{*}{$f_{\rm esc}^{\rm LyC}$ model} & \multirow{2}{*}{$M_{\rm UV, lim}$} & \multirow{2}{*}{$\log_{10}(\dot{n}_{\rm ion}/\rm s^{-1} Mpc^{-3})$} & \multirow{2}{*}{$\sigma_{\log_{10}\dot{n}_{\rm ion}}$} & \multicolumn{3}{c}{$C_{\rm HII, rec} \quad (\dot{Q}_{\rm HII}=0)$} \\ 
    & & & & $Q_{\rm HII} = 1.00$ & $Q_{\rm HII} = 0.85$ & $Q_{\rm HII} = 0.70$ \\ 
    \hline 
    \multirow{3}{*}{$10\%$} & $-17.0$ & $50.61^{+0.05}_{-0.05}$ & $+0.19-0.18$ & $3.4^{+0.4}_{-0.4}$ & $4.0^{+0.5}_{-0.4}$ & $4.9^{+0.6}_{-0.5}$ \\ 
     & $-15.5$ & $50.82^{+0.09}_{-0.08}$ & $+0.20-0.18$ & $5.5^{+1.3}_{-1.0}$ & $6.4^{+1.6}_{-1.1}$ & $7.8^{+1.9}_{-1.4}$ \\ 
     & $-13.5$ & $51.03^{+0.17}_{-0.14}$ & $+0.20-0.18$ & $8.8^{+4.3}_{-2.5}$ & $10.4^{+5.0}_{-2.9}$ & $12.6^{+6.1}_{-3.5}$ \\ 
    \hline 
    \multirow{3}{*}{Chisholm+22} & $-17.0$ & $50.31^{+0.07}_{-0.06}$ & $+0.39-0.36$ & $1.7^{+0.3}_{-0.2}$ & $2.0^{+0.3}_{-0.3}$ & $2.4^{+0.4}_{-0.3}$ \\ 
     & $-15.5$ & $50.58^{+0.13}_{-0.11}$ & $+0.40-0.37$ & $3.1^{+1.0}_{-0.6}$ & $3.7^{+1.2}_{-0.8}$ & $4.5^{+1.5}_{-0.9}$ \\ 
     & $-13.5$ & $50.87^{+0.23}_{-0.20}$ & $+0.40-0.39$ & $6.2^{+4.1}_{-2.1}$ & $7.3^{+4.8}_{-2.5}$ & $8.9^{+5.8}_{-3.1}$ \\ 
    \botrule
    \end{tabular}
    \caption{\ndotion\ and \cHII\ values determined as the summation of star forming and smouldering \ndotion$-$\MUV\ curves from  \autoref{fig:ndot_ion_integrand} integrated to UVLF turnover magnitudes, $M_{\rm UV, lim}=\{-17.0, -15.5, -13.5\}$ for our fixed $f_{\rm esc}^{\rm LyC}=0.1$ and \citet{Chisholm2022} derived $f_{\rm esc}^{\rm LyC}(M_{\rm UV})$ models. The median scatter in \ndotion\ is given by $\sigma_{\log_{10}\dot{n}_{\rm ion}}$. \cHII\ values are computed assuming three different \ion{H}{2} volume filling factors $Q_{\rm HII}=\{1.00, 0.85, 0.70\}$ consistent with more direct measurements of the neutral fraction at $z=6$.}
    \label{tab:ndot_ion_CHII_values}
\end{table*}

\autoref{tab:ndot_ion_CHII_values} additionally shows the scatter in \ndotion\ represented by $\sigma_{\log_{10}\dot{n}_{\rm ion}}$. While it is likely that the patchiness of reionization will contribute non-negligibly, this scatter is likely instead dominated by the uncertainty in \citet{Calzetti2000} dust attenuation estimates determined by our SED fitting procedure. Since it remains challenging to quantify the impact of dust on our observed scatter, we do not attempt to decouple the intrinsic scatter in IGM \ion{H}{2} clumping from that induced by our observational methodology, and hence do not quote the observed scatter on the measurement of $C_{\rm HII, rec}$. We note that our $\sigma_{\log_{10}\dot{n}_{\rm ion}}$ measurements for the \citet{Chisholm2022} \fesc\ model do not include the scatter in the conversion between $\beta_{\rm UV}$ and \fesc\ within the LzLCS sample, which would add to this observed scatter.

\subsection{Modelling caveats}

\subsubsection{Self-shielding ``sinks'' for ionizing photons at the end of the EoR}

The reionization ODE given by \autoref{eq:reionization_ode} balances the rates of hydrogen-only ionization and recombination assuming that all emitted photons escaping the ISM of their host galaxy also escape the CGM and ionize HI residing in the IGM. This approximation breaks down rapidly towards the end of reionization when the \textit{average} mean free path of ionizing photons, $\langle\lambda_{\rm mfp}\rangle$, rapidly increases during the epoch of ionized bubble overlap. 

The source of reionization ``sinks'' is dominated by Lyman-Limit Systems (LLSs) with HI column densities $N_{\rm HI}\gtrsim10^{17}\rm cm^{-2}$ due to their relative abundance over damped Lyman alpha systems \citep{Furlanetto2009, Haardt2012, Sobacchi2014}. These LLSs are self-shielded, dominating the mean free path of ionizing photons; the mean opacity ($\kappa_{\nu}\propto\lambda_{\rm mfp,\nu}^{-1}$) of photons at the Lyman-limit ($\nu=\nu_{\rm L}=3.29\times10^{15}~\rm Hz$) is thus given by $\langle\kappa_{\nu_{\rm L}}\rangle = \langle\kappa_{\nu_{\rm L}}^{\rm LLS}\rangle + \langle\kappa_{\nu_{\rm L}}^{\rm IGM}\rangle$. 

This can be used to modify the source term of the reionization ODE in \autoref{eq:reionization_ode} by a factor $(1 + \langle\kappa_{\nu_{\rm L}}^{\rm LLS}\rangle / \langle\kappa_{\nu_{\rm L}}^{\rm IGM}\rangle)$ in the denominator, suppressing our inferred $C_{\rm HII, rec}$ by the same factor \citep{Madau2017}. Since $\langle\kappa_{\nu_{\rm L}}^{\rm IGM}\rangle\propto\langle x_{\rm HI}\rangle$, this factor only becomes important during the end stages of the EoR when the HI content is low; it remains critically dependent on the residual HI content within ionized bubbles, estimated to be $\langle x_{\rm HI}\rangle\sim10^{-4}$ at $z\simeq6$ by observations of the Gunn-Peterson trough in SDSS quasar spectra by \citet{Fan2006}. The $\langle\kappa_{\nu_{\rm L}}^{\rm LLS}\rangle$ term is dependent on the number density of LLSs, obtained by stacked quasar spectra at $z\lesssim5.5$ \citep[e.g.][]{Prochaska2009,O'Meara2013,Worseck2014}. 

Combining these results yields $(1 + \langle\kappa_{\nu_{\rm L}}^{\rm LLS}\rangle / \langle\kappa_{\nu_{\rm L}}^{\rm IGM}\rangle)\sim3$ at $z\simeq6$, which increases with time reaching $\sim2.8$ at $z\simeq5$ and $\rightarrow\infty$ at $z\simeq4.3$. This indicates that our inferred values for $C_{\rm HII, rec}$ may be a factor $\sim3$ overestimated; a faint UVLF turnover limit of $M_{\rm UV, lim}\gtrsim-12.3$ would thus be required to favour the \ion{H}{2} clumping prescription in \citet{Davies2024} over the \citet{Shull2012} model assuming $f_{\rm esc}^{\rm LyC}=0.1$ and total stable reionization. We note that in the case of later reionization models with $Q_{\rm HII}(z=6)\neq1$, the overlap of ionized bubbles is not sufficient to provide a notable difference in the estimated value of $C_{\rm HII, rec}$.



\subsubsection{The topology of reionization}
\label{sec:reionization_topology}

It is important to note that as we determine the clumping factor from the timescale required to equate the photoionization and recombination rates in a stably reionized Universe with given $Q_{\rm HII}$, we are specifically inferring the \textit{recombination-averaged global} hydrogen clumping factor, defined mathematically as 
\begin{equation}
    C_{\rm HII, rec} = \frac{\langle n_{\rm H}^2(1+x_{\rm He})x_{\rm HII}\alpha_{\rm rec}(T_e)\rangle}{\langle n_{\rm H}\rangle^2\langle (1 + x_{\rm He})\rangle\langle x_{\rm HII}\rangle\langle \alpha_{\rm rec}(T_e)\rangle}.
    \label{eq:cHII_definition}
\end{equation}
Crucially, this is not equivalent to the physical clumping of the gas distribution of hydrogen, $C_{\rm H}=\langle n_{\rm H}^2\rangle/\langle n_{\rm H}\rangle^2$, since it dependent upon both the inhomogeneous temperature distribution and the topology of reionization for both hydrogen and, to a lesser extent, helium.

The topology of reionization, remains critically important for inferences regarding the clumping of gas. Two plausible scenarios exist, namely ``inside-out'', where the Universe reionizes hierarchically from the most to the least-dense regions, and ``outside-in'', which is exactly the opposite. Should reionization indeed be ``inside-out'', the velocity of ionization fronts will be increased towards the end of the EoR as the less dense voids finish filling in, making our $\mathrm{d}Q_{\rm HII}/\mathrm{d}t=0$ assumption made between \autoref{eq:reionization_ode} and \autoref{eq:stable_reionization} increasingly invalid. While cosmological hydrodynamical simulations favour this ``inside-out'' reionization \citep[e.g.][]{Pawlik2009,Finlator2011,Shull2012,McQuinn2011,Pawlik2015}, the pioneering work of \citet{Miralda-Escude2000} strongly favours the ``outside in'' model where the clumping falls below unity (i.e. the gas recombines more efficiently than in an homogeneous IGM) when all gas below an overdensity $\Delta=\rho/\bar{\rho}\sim10$ at $z\simeq6$ is reionized.

The $C_{\rm HII, rec}$ weighting in \autoref{eq:cHII_definition} is made more complex by the $\alpha_{\rm rec}(T_e)\appropto T_e^{-0.7}$ temperature dependence. Since overdense regions are on average hotter \citep{Hui1997}, recombination timescales should be longer in these regions, favouring ``inside-out'' reionization and observed clumping factors. In contrast, the Jeans length temperature dependence, $\lambda_{\rm J}\propto T^{1/2}$, means that the smallest mass clumps are ``smoothed'' out as the temperature increases, reducing $C_{\rm HII, rec}$ in a process known as photoevaporation. Nevertheless, we note that our elevated \cHII\ measurements could in part be due to reionization being more ``inside-out'' than described by simulations.

\subsubsection{Considering the reionization history}

The requirement of reionization to progress over time means that our $\mathrm{d}Q_{\mathrm{HII}}/\mathrm{d}t=0$ assumption made in \autoref{eq:stable_reionization} is incorrect by construction. To evaluate the impact of this on our inferred \cHII\ values, we compute the full integral in \autoref{eq:reionization_ode} to calculate $Q_{\rm HII}$ at $z=6$ on the same $\{C_{\rm HII, rec},M_{\rm UV, lim}\}$ grid as plotted in \autoref{fig:cHII_MUVlim} numerically using {\tt{scipy.integrate.solve\_ivp}}. This takes into account the previous reionization history; we assume a starting redshift, $z_{\rm start}=15$, redshift independent \cHII, constant IGM electron temperature $T_e=10^4\,\rm K$. Three runs are performed assuming a different \ndotion\ redshift dependence, \dndotdz\,$\in[0.0, -0.2, -0.4]$, consistent with the evolution presented in \citet{Bouwens2015-EoR} and \citet{Finkelstein2019}.

The $Q_{\rm HII}=\{1.00, 0.85, 0.70\}$ contours from the numerical integration are compared to the $\mathrm{d}Q_{\rm HII}/\mathrm{d}t=0$ fiducial values to produce a correction term, $\Delta C_{\rm HII, rec}= C_{\rm HII,rec}^{\,\dot{Q}_{\rm HII}\neq0} - C_{\rm HII,rec}^{\,\dot{Q}_{\rm HII}=0}$, to account for the reionization history, given by
\begin{equation}
    \Delta C_{\rm HII, rec}\simeq 19.1 \times \frac{\mathrm{d}\log_{10}\dot{n}_{\rm ion}}{\mathrm{d}z} -2.5.
    \label{eq:clumping_factor_evolution_correction}
\end{equation}
Our \dQdtzero\ assumption is thus only valid for the physically unmotivated \dndotdz\,$\simeq0.13$. The observed \dndotdz\,$\simeq-0.2$ from \citet{Bouwens2015-EoR} and \dndotdz\,$\simeq0.0$ from \citet{Finkelstein2019} would suggest $\Delta C_{\rm HII,rec}\simeq -6.3$ and $\Delta C_{\rm HII,rec}\simeq -2.5$ respectively.

It is noteworthy that the clumping factor correction in \autoref{eq:clumping_factor_evolution_correction} holds only in the \cHII\,$>0$ regime where enough ionizing photons are released into the IGM cumulatively during the entire history of reionization by galaxies to produce a Universe with neutral content $Q_{\rm HII}$ \textit{without recombinations}. A comprehensive comparison of early and late reionization models is presented in \citet{Cain2025}.

\subsubsection{The contribution by AGN}
\label{sec:AGN_contribution}

It is known that AGN provide the dominant source of ionizing photons for the $\rm HeII$ reionization at $z\sim4$ and a non-negligible contribution to HI reionization \citep[e.g.][]{Yung2021, Asthana2024, Dayal2025}. In this work we have implicitly assumed that galaxies contribute $100\%$ to hydrogen reionization; any additional contribution from AGN will increase the total $\dot{n}_{\rm ion}^{\rm tot}=\dot{n}_{\rm ion}^{\rm gal} + \dot{n}_{\rm ion}^{\rm AGN}$ and thus the inferred \cHII\ required for stable reionization. Our \cHII\ measurements should therefore be taken as lower limits as a result of this unknown AGN contribution.

By integrating the AGN luminosity function to determine the emissivity of LyC sources \citep[see e.g.][]{Haardt2012, Madau2015}, several studies have suggested that AGN contribute $\lesssim10\%$ of the ionizing photon budget \citep[e.g.][]{Kulkarni2019}, backed up by results from simulations, including DELPHI \citep[][]{Dayal2020,Mauerhofer2025} and OBELISK \citep[][]{Trebitsch2021}. More recent studies, however, highlight that either an increased AGN $f_{\rm esc}^{\rm LyC}\gtrsim60\%$ \citep{Madau2024, D'Silva2025} or independent modelling of faint, Type I, unobscured AGN \citep{Singha2025} could increase the contribution of AGN to more than this fiducial $10\%$.

Integrating the \ndotion$-z$ relations from \citet{Kulkarni2019} and \citet{D'Silva2025} over the redshift range used in this work ($5.6<z<6.5$) yields $\log_{10}(\dot{n}_{\rm ion}^{\rm AGN}/\rm s^{-1}Mpc^{-3})=49.27\pm0.50$ and $\log_{10}(\dot{n}_{\rm ion}^{\rm AGN}/\rm s^{-1}Mpc^{-3})=50.53\pm0.10~\rm s^{-1}Mpc^{-3}$ respectively. Comparing these to our $f_{\rm esc}^{\rm LyC}=0.1$, $M_{\rm UV, lim}=-13.5$ galaxy $\dot{n}_{\rm ion}^{\rm gal}$ presented in \autoref{tab:ndot_ion_CHII_values} gives AGN which provide $2\%$ \citep{Kulkarni2019} and $24\%$ \citep{D'Silva2025} of the total ionizing photon budget. An increase in \cHII\ by factors of $1.02$ and $1.32$ respectively are required to increase the number of recombinations to stabilize reionization against the increased emissivity. Should we instead relax our UVLF turnover limit to $M_{\rm UV, lim}=-17.0$, the AGN contributions (\cHII\ multiplicative factors) changes to $4\%$ ($1.05$) for \citet{Kulkarni2019} and $45\%$ ($1.83$) for \citet{D'Silva2025}. It is worth noting that these fractions are lower limits due to the unknown contribution from obscured Type II AGN.

\section{Conclusions} \label{sec:conclusions}

In summary, we present indirect \ion{H}{2} clumping factor results at $5.6<z<6.5$ using a sample of 1721 candidate Lyman Break Galaxies (LBGs) at $5.6<z<6.5$ across $550$\arcmin$^2$ of unmasked sky area across the JADES (GOODS-North/South), PRIMER (UDS/COSMOS), and NEP surveys collated in {\tt EPOCHS-v2} (Austin et al. in prep). We use \bagpipes\ with \bpass\ v2.2.1 SPS templates and a ``continuity bursty'' SFH with an additional $3-10~\rm Myr$ age bin to compute \xiionzero\ and \Ndotionzero\ for our sample, which is split by star formation burstiness, $\rm SFR_{10}/SFR_{100}$ into star forming and smouldering subsamples. We perform a 90\% stellar mass cut, with limits measured between $\log_{10}(M_{\star, \rm lim}/\mathrm{M}_{\odot})\in[8.1, 9.2]$ using the {\tt JAGUAR} mock catalogue.

We measure a flatter
\begin{equation*}
    \beta_{\rm UV}=(-0.040\pm0.022)M_{\rm UV}-2.88^{+0.43}_{-0.44},
\end{equation*}
than measured previously with \hst\ at $z\simeq6$ \citep[e.g.][]{Bouwens2012}, with $0.27~\rm dex$ scatter. This highlights the new population of intrinsically bright, dust free sources which may leak copious amounts of LyC photons.

The ionizing photon production efficiency, \xiionzero, is computed from the \ha\ recombination line flux. Our \bagpipes\ models reach a maximum $\log_{10}(\xi_{\rm ion,0}/\rm Hz\,erg^{-1})=25.75$, highlighting the limitations of SED fitting which cannot reproduce the extreme values $\log_{10}(\xi_{\rm ion,0}/\rm Hz\,erg^{-1})\gtrsim25.8$ observed spectroscopically with JWST/NIRSpec. We find no trend of \xiionzero\ with \MUV, measuring
\begin{equation*}
    \log_{10}(\xi_{\rm ion, 0}/\rm Hz\,erg^{-1}) = -0.006^{+0.019}_{-0.017}~M_{\rm UV} + 25.05^{+0.39}_{-0.34}
\end{equation*}
for the full sample, consistent with canonical $\xi_{\rm ion, 0}\simeq10^{25.2}~\rm Hz~erg^{-1}$ values at $z\simeq6$. Our SFG subsample produces results $0.1-0.15~\rm dex$ higher. The gradient of this relation differs from recent observational results suggesting either an increasing \citep{Simmonds2024a, Begley2025b, Papovich2025, Llerena2025a} or decreasing \citep{Simmonds2024b, Begley2025a, Pahl2025a, Pahl2025b} \xiionzero\ for fainter galaxies.

We calculate the $z\simeq6$ UVLF for our combined, star forming, and smouldering samples, complete down to $M_{\rm UV}\simeq-18$ in the deepest regions of JADES-GOODS-South. We find results consistent with \hst\ HFF and CANDELS studies, although the SFG subsample has a steeper faint-end slope  ($\alpha_{\rm SF}=-2.2\pm0.2$) than our smouldering sources ($\alpha_{\rm sm}=-1.7\pm0.2$).

Combining our measured $\dot{N}_{\rm ion,0}-M_{\rm UV}$ relation with our Schechter fits to the $z\simeq6$ UVLFs, we compute the total contribution to reionization, \ndotion. Assuming the UVLF turnover occurs at $M_{\rm UV}=-17$, we find $\log_{10}(\dot{n}_{\rm ion}/\rm s^{-1}Mpc^{-3})=50.31^{+0.07}_{-0.06}$ adopting the \MUV\ dependent \fesc\ relation derived from our $\beta_{\rm UV}$ measurements, and $\log_{10}(\dot{n}_{\rm ion}/\rm s^{-1}Mpc^{-3})=50.61^{+0.05}_{-0.05}$ with fixed $f_{\rm esc}^{\rm LyC}=10\%$. These results are consistent with previous JWST studies by \citet{Simmonds2024b}.

Assuming stable reionization ($\mathrm{d}Q_{\rm HII}/\mathrm{d}t=0$), the UVLF turnover constraints from GLIMPSE (i.e. $M_{\rm UV, lim}\simeq-13.5$, Goolsby et al. in prep.) suggest a clumping factor \cHII\,$=6.2^{+4.1}_{-2.1}$ is required to fully reionize the Universe by $z\simeq6$. If we instead assume a constant \fesc\,$=10\%$, we measure \cHII\,$=8.8^{+4.3}_{-2.5}$. Clumping factors of this magnitude resolve the ionizing photon budget crisis outlined by \citet{Munoz2024}. 

Our fiducial results favor the \citet{Davies2024} clumping factor measurements derived observationally from mean free path measurements in stacked quasar spectra over the more traditional \citet{Shull2012} relation at $z\simeq6$ for UVLF turnover magnitudes $M_{\rm UV, lim}>-12.7$ ($M_{\rm UV, lim}>-14.2$ for \fesc\,$=10\%$). While these are intriguing, we note that our results are sensitive to \fesc, $x_{\rm HI}$, the contribution of self-shielded LLSs within ionized bubbles, the topology and history of reionization, and the contribution from AGN. None of these factors are well constrained and can lead to dramatically different \cHII\ results.




Nevertheless, our methodology paves the way for future \cHII\ constraints pre-overlap where more direct quasar analyses become impossible due to the saturation of the \lya\ forest. Future \xiion\ measurements from JWST/NIRCam at $6.5<z<7.6$ utilizing the \fion{O}{3}+\hbeta\ optical emission line complex will constrain the reionization history and allow for a more robust measurement of \cHII. 

\begin{acknowledgments}

\vspace{10pt}

We acknowledge support from the ERC Advanced Investigator Grant EPOCHS (788113), as well as two studentships from STFC for DA and TH. LW acknowledges funding from the Faculty of Science \& Engineering at the University of Manchester. RAW, SHC, and RAJ acknowledge support from NASA JWST Interdisciplinary Scientist grants NAG5 12460, NNX14AN10G and 80NSSC18K0200 from GSFC. 

This work is based on observations made with the NASA/ESA \textit{Hubble Space Telescope} (HST) and NASA/ESA/CSA \textit{James Webb Space Telescope} (JWST) obtained from the \texttt{Mikulski Archive for Space Telescopes} (\texttt{MAST}) at the \textit{Space Telescope Science Institute} (STScI), which is operated by the Association of Universities for Research in Astronomy, Inc., under NASA contract NAS 5-03127 for JWST, and NAS 5–26555 for HST. Some of the data products presented herein were retrieved from the Dawn JWST Archive (DJA). DJA is an initiative of the Cosmic Dawn Center, which is funded by the Danish National Research Foundation under grant No. 140. The authors thank all involved in the construction and operations of the telescope as well as those who designed and executed these observations, their number are too large to list here and without each of their continued efforts such work would not be possible. 


The authors thank Anthony Holloway and Sotirios Sanidas for their providing their expertise in high performance computing and other IT support throughout this work. 

\end{acknowledgments}

\begin{contribution}

We list here the roles and contributions of the authors according to the Contributor Roles Taxonomy (CRediT)\footnote{\url{https://credit.niso.org}}. \\
\textbf{Duncan A. Austin}: Conceptualization, Data curation, Formal analysis, Investigation, Methodology, Project Administration, Resources, Software, Validation, Visualization, Writing - original draft, Writing - review \& editing \\
\textbf{Thomas Harvey}: Software, Writing - review \& editing \\
\textbf{Christopher J. Conselice}: Funding Acquisition, Project Administration, Supervision, Writing - review \& editing \\
\textbf{Nathan J. Adams}: Data curation, Resources \\
\textbf{Vadim Rusakov}: Investigation, Resources, Software \\
\textbf{Sophie L. Newman, Louise T. C. Seeyave}: Data curation, Writing - review \& editing \\
\textbf{Massimo Ricotti:} Methodology, Writing - review \& editing \\
\textbf{Caio Goolsby, James Trussler, Brenda Frye, Norman A. Grogin, Rolf A. Jansen, Anton M. Koekemoer, Michael Rutkowski, Rogier A. Windhorst:} Writing - review \& editing \\

\end{contribution}

\section*{Data Availability}

Photometric imaging and catalogues will be made available upon reasonable request to the corresponding author.

\facilities{HST(ACS-WFC), JWST(NIRCam)}

\software{
astropy \citep{2013A&A...558A..33A,2018AJ....156..123A,2022ApJ...935..167A}, 
\cloudy\ \citep{Ferland2013}, \texttt{PyMC} \citep{Salvatier2015_pymc},
\sextractor\ \citep{Bertin1996-SExtractor}
}


\newpage

\appendix

\section{Power law fitting results}
\label{sec:power_law_params}

Throughout this paper we have fit power laws, 
\begin{equation}
    Y=\frac{\mathrm{d}Y}{\mathrm{d}M_{\rm UV}}M_{\rm UV} + Y_0, 
    \label{eq:power_law}
\end{equation}
to produce scaling relations for $Y=\{\beta_{\rm UV}, \xi_{\rm ion, 0}, \dot{N}_{\rm ion,0}\}$ in terms of intrinsic UV magnitude, \MUV. The scatter in our relation is measured in general in terms of some normalization, $\sigma_c$, some $Y$-dependent scatter, $\sigma_Y$, and some \MUV-dependent scatter, $\sigma_{M_{\rm UV}}$, which can be modelled as
\begin{equation}
    \sigma(M_{\rm UV}, Y)=\lvert \sigma_c + Y\sigma_Y + M_{\rm UV}\sigma_{M_{\rm UV}}\rvert.
    \label{eq:power_law_scatter}
\end{equation}
For $Y=\beta_{\rm UV}$ our errors are Gaussian, hence we include only $\sigma_c$ dependence, whereas the heteroscedasticity in \ha\ luminosity measurements prompts us to include $\sigma_Y$ dependence for $Y=\xi_{\rm ion,0}$, and both $\sigma_Y$ and $\sigma_{M_{\rm UV}}$ dependence for $Y=\dot{N}_{\rm ion,0}$. In each case we perform Bayesian MCMC fitting using \autoref{eq:power_law} and \autoref{eq:power_law_scatter} adopting wide uniform priors using the {\tt emcee} python package \citep{Foreman-Mackey2013}. We restrict the bounds of the scatter to exclude the possibility of bimodal solutions. Results for our combined, SFG and smouldering samples at $z\simeq6$ are presented in \autoref{tab:power_law_params}.

\begin{table}[h]
   \centering
   \caption{Power law fitting posteriors for our $\beta_{\rm UV}-M_{\rm UV}$ (\autoref{fig:EPOCHS_v2_beta_MUV}), \xiionzero$-M_{\rm UV}$ (\autoref{fig:EPOCHS_v2_xi_ion_MUV}), and \Ndotionzero$-M_{\rm UV}$ (\autoref{fig:Ndot_ion_MUV}) relations. Here we parametrize the fit as $Y=(\mathrm{d}Y/\mathrm{d}M_{\rm UV})M_{\rm UV} + Y_0$ and the scatter as $\sigma(M_{\rm UV}, Y) = \lvert\sigma_{\rm c} + Y\sigma_Y + M_{\rm UV}~\sigma_{\rm MUV}\rvert$}
   \setlength{\tabcolsep}{2pt}
   \begin{tabular}{cc|cc|ccc}
   \toprule
   Independent variable, $Y$ & Sample & $\mathrm{d}Y/\mathrm{d}M_{\rm UV}$ & $Y_0$ &  $\sigma_{\rm c}$ & $\sigma_Y$ & $\sigma_{M_{\rm UV}}$\\
   \hline
   \multirow{3}{*}{$\beta$} & Total & $-0.040\pm0.022$ & $-2.88^{+0.43}_{-0.44}$ & $0.27\pm0.02$ & $-$ & $-$ \\
   & Star-forming & $-0.007^{+0.035}_{-0.031}$ & $-2.25^{+0.70}_{-0.61}$ & $0.25^{+0.05}_{-0.04}$ & $-$ & $-$ \\
   & Smouldering & $-0.051^{+0.025}_{-0.026}$ & $-3.07^{+0.49}_{-0.50}$ & $0.28\pm0.03$ & $-$ & $-$ \\
   \hline
   \multirow{3}{*}{\xiionzero\ } & Total & $-0.006^{+0.019}_{-0.017}$ & $25.05^{+0.39}_{-0.34}$ & $18.98^{+1.13}_{-1.07}$ & $-0.754^{+0.019}_{-0.017}$ & $-$ \\
   & Star-forming & $0.011^{+0.013}_{-0.015}$ & $25.54^{+0.25}_{-0.29}$ & $17.74^{+3.45}_{-3.04}$ & $-0.701^{+0.120}_{-0.136}$ & $-$ \\
   & Smouldering & $-0.001\pm0.015$ & $24.43^{+0.31}_{-0.30}$ & $13.49^{+1.89}_{-1.83}$ & $-0.556^{+0.074}_{-0.078}$ & $-$ \\
   \hline
    \multirow{3}{*}{\Ndotionzero\ } & Total & $-0.542^{+0.047}_{-0.039}$ & $43.20^{+0.96}_{-0.78}$ & $34.26^{+2.30}_{-2.35}$ & $-0.800^{+0.046}_{-0.044}$ & $-0.444^{+0.056}_{-0.051}$ \\
   & Star-forming & $-0.417^{+0.026}_{-0.027}$ & $46.02^{+0.52}_{-0.55}$ & $40.94^{+4.41}_{-4.17}$ & $-0.889^{+0.090}_{-0.096}$ & $-0.370^{+0.048}_{-0.054}$\\
   & Smouldering & $-0.391^{+0.049}_{-0.041}$ & $45.43^{+0.92}_{-0.78}$ & $25.07^{+3.76}_{-3.73}$ & $-0.561^{+0.075}_{-0.077}$ & $-0.235^{+0.059}_{-0.058}$\\
   \botrule
   \end{tabular}
   \label{tab:power_law_params}
\end{table}


\section{Determining the \texorpdfstring{\xiionzero\ sensitivity limit}{sensitivity limit of ionizing photon production efficiency measurements} at \texorpdfstring{$5.6<\lowercase{z}<6.5$}{5.6<z<6.5}}
\label{sec:xi_ion_sensitivity}

\begin{figure}[h!]
    \centering
    \includegraphics[width=0.7\linewidth]{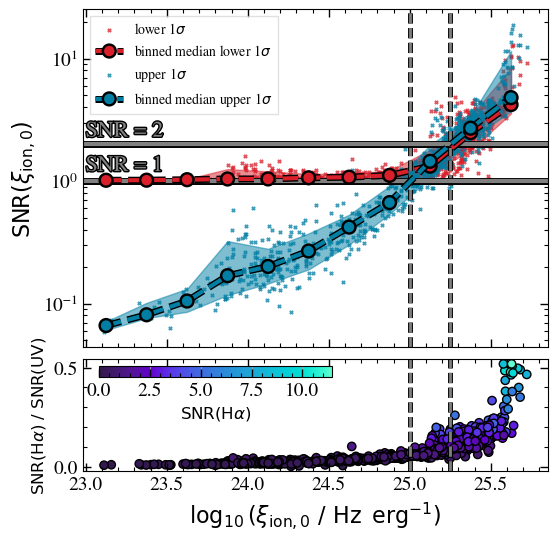}
    \caption{Limiting sensitivity of \xiionzero\ measurements at $5.6<z<6.5$. Vertical dashed lines show the approximate limiting $1\sigma$ ($\log_{10}(\xi_{\rm ion,0}/\rm Hz~erg^{-1})\simeq25.0$) and $2\sigma$ ($\log_{10}(\xi_{\rm ion,0}/\rm Hz~erg^{-1})\simeq25.25$) sensitivities. SNRs are calculated using the lower (red) and upper (blue) $1\sigma$ errors, where the large circular points and shaded regions show the median and $16^{\rm th}-84^{\rm th}$ percentiles in $0.25~\rm dex$ bins. The ratio of \ha\ to UV continuum SNRs are shown in the lower inset.}
    \label{fig:xi_ion_sensitivity}
\end{figure}

From the right hand panel in \autoref{fig:bagpipes_xi_ion}, it is apparent that there exists some sensitivity limit below which we cannot precisely measure \xiionzero\ from the photometric data. We investigate this limit by producing mock photometry at $5.6<z<6.5$ from the first realization of the {\tt JAGUAR} catalogues introduced in \autoref{sec:EPOCHS_v2_sample_completeness} using the same mock depths and filterset as for the mock photometry from the spectra described above. Following this, we run the catalogue through our selection criteria; we are left with 2,064 sources which are then run through \bagpipes\ to calculate \xiionzero\ values. Results are shown in \autoref{fig:xi_ion_sensitivity}, where we can see that $1\sigma$ and $2\sigma$ sensitivities reside at $\log_{10}(\xi_{\rm ion,0}/\rm Hz~erg^{-1})\simeq25.0$ and $\log_{10}(\xi_{\rm ion,0}/\rm Hz~erg^{-1})\simeq25.25$ respectively.

Taking the more conservative $2\sigma$ limit, we note that anything below $\log_{10}(\xi_{\rm ion,0}/\rm Hz~erg^{-1})\simeq25.25$ can be considered an upper limit (as opposed to a lower limit due as a consequence of the larger UV continuum SNRs compared to the corresponding \ha\ line SNRs). Further studies incorporating shallower but more sensitive narrow or medium bands at $\gtrsim4\mu\rm m$, for instance with the JWST Emission Line Survey \citep[JELS; see e.g.][]{Duncan2025,Pirie2025}, will improve the \xiionzero\ sensitivity of our measurements.

\newpage

\section{The efficiency of our burstiness selection criteria with spectroscopy}
\label{sec:burstiness_selection_efficiency}

\begin{figure}[h!]
    \centering
    \includegraphics[width=0.7\linewidth]{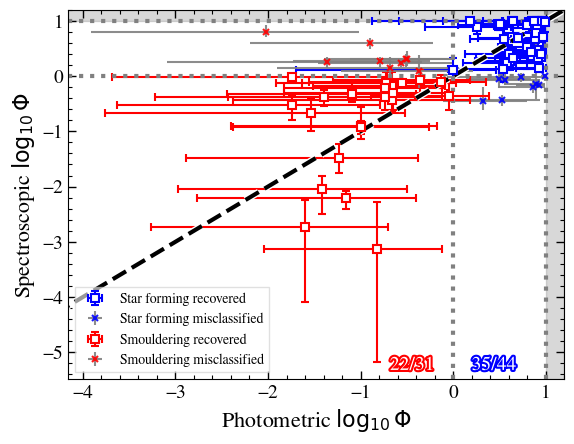}
    \caption{Comparison of photometric and spectroscopic estimates of star formation burstiness from 71 NIRSpec PRISM/CLEAR spectra at $5.6<z<6.5$ and via photometric and spectroscopic \bagpipes\ SED fitting. The purity of our subsamples is estimated to be $\simeq80\%$ (star forming) and $\simeq70\%$ (smouldering).}
    \label{fig:phot_spec_burstiness_comparison}
\end{figure}

To estimate the spectroscopic burstiness values, we fit 75 NIRSpec PRISM/CLEAR spectra at $5.6<z<6.5$ collated from the DJA v4.2 (explained in more detail in \autoref{sec:xi_ion_spec}) directly with \bagpipes\ at their fixed DJA catalogued redshifts. The same fiducial prior setup as with the photometry is used and we adopt the default spectral resolution curve distributed by STScI. A comparison of photometric and spectroscopic burstiness estimates are presented in \autoref{fig:phot_spec_burstiness_comparison}. 

We find that $35/44$ ($\simeq80\%$) of photometrically selected star forming sources (with $\Phi>1$) are correctly identified, compared to only $22/31$ ($\simeq70\%$) of photometrically smouldering sources. This technique assumes, as always, that our SFH prescription adequately describes our sources, meaning these values are more likely the best-case scenario. Nevertheless, we conclude that our star forming subsample has marginally more purity, although both subsamples remain relatively interloper-free.

\section{The impact of \texorpdfstring{\fion{N}{2}}{[NII]} on measurements of \texorpdfstring{\xiionzero}{ionizing photon production efficiency}}
\label{sec:nii_halpha}

Our conclusions in this work are dependent in part on the measured $\rm H \alpha$ line fluxes, which are determined from the photometric boost in the F444W bandpass compared to the continuum which is traced by F410M. It is important, therefore, to determine the contribution to this F410M$-$F444W colour by nearby emission lines other than the \ha\ recombination line. Most notable of which are the \fion{N}{2}$_{\lambda6548}$ and \fion{N}{2}$_{\lambda6584}$ collisionally excited lines arising from $\rm N^{+}$.

\begin{figure}[h!]
    \centering
    \includegraphics[width=0.7\linewidth]{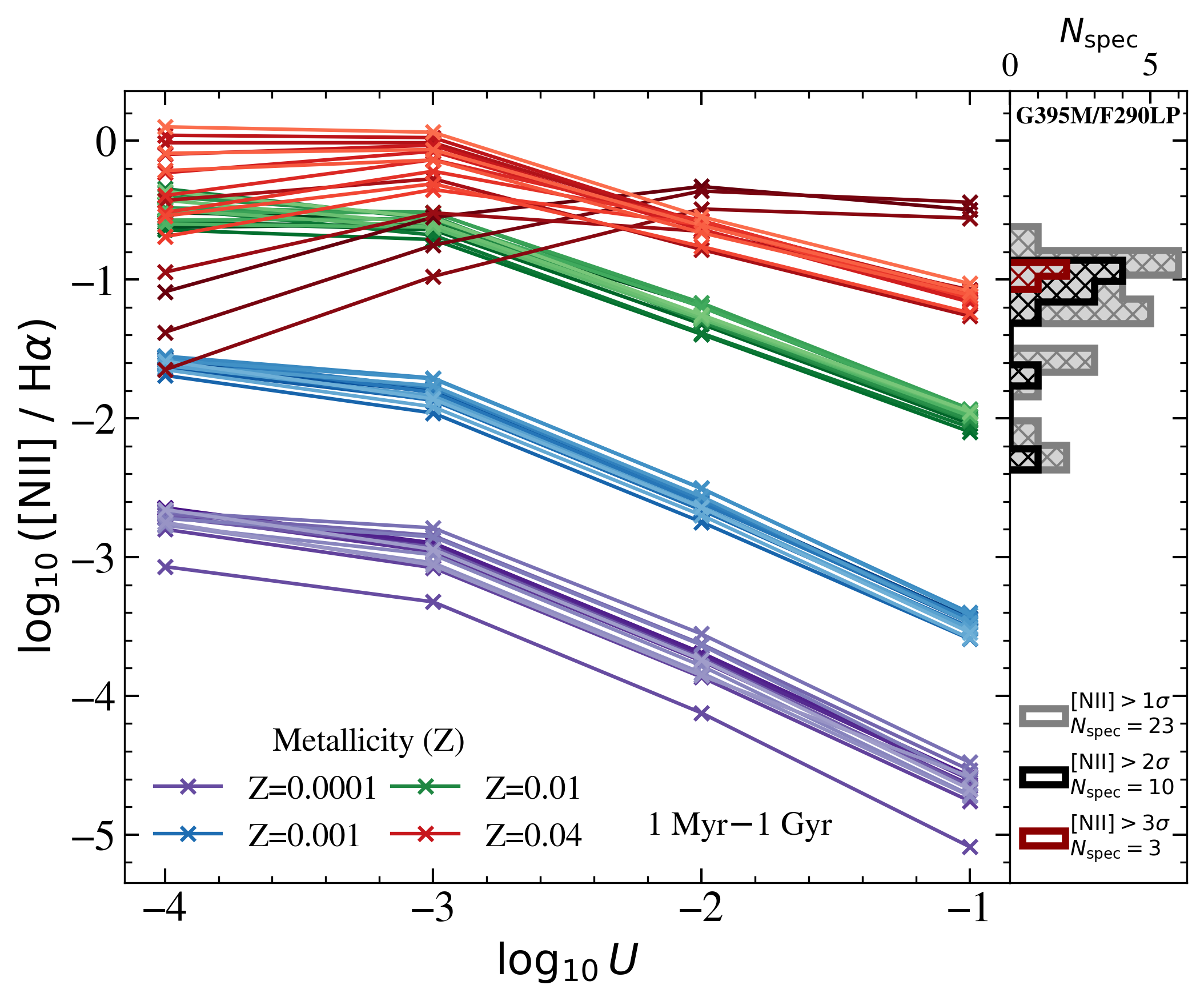}
    \caption{\fion{N}{2}/\ha\ ratio as a function of ionization parameter $\log_{10}U\in[-4, -1]$ for metallicities $Z\in[10^{-4}, 0.04]$ and ages $1~\rm Myr-1~\rm Gyr$ (with darker lines showing older models) computed using \textsc{synthesizer} \citep{Lovell2025, Roper2025}. These models use the \bpass-2.2.1 SPS binary-star models run through \cloudy-v23.01 assuming a \citet{Chabrier2003} IMF with mass limits $M_{\star}=[0.1,300]\rm M_{\odot}$ following the prescription in \citet{Newman2025}. The right hand panel shows observational results from JWST/NIRSpec G395M/F290LP spectra at $5.6<z<6.5$, $\{23,10,3\}$ of which have both \fion{N}{2}$_{\lambda6548}$ and \fion{N}{2}$_{\lambda6584}$ detected at $\{1,2,3\}\sigma$.}
    \label{fig:NII_Halpha_ratio}
\end{figure}

We calculate the \fion{N}{2}/\ha\ ratio (\fion{N}{2}~$\equiv$~\fion{N}{2}$_{\lambda6548}+$\fion{N}{2}$_{\lambda6584}$) from mock galaxies produced by the \textsc{synthesizer} python package \citep{Lovell2025, Roper2025} as a function of ionization parameter $\log U\in[-4,-1]$ using the grids computed in \citet{Newman2025}. These models use the \bpass-2.2.1 SPS binary star models run through \cloudy-v23.01 assuming a \citet{Chabrier2003} IMF with mass limits $M_{\star}=[0.1,300]~\rm M_{\odot}$, and are computed at fixed metallicities $\log_{10}Z\in[10^{-4}, 0.04]$ and ages $1~\rm Myr-1~\rm Gyr$. Results are shown in the main panel of \autoref{fig:NII_Halpha_ratio}. There is a strong correlation such that high metallicity systems have larger \fion{N}{2}/\ha. \fion{N}{2}/\ha falls with increasing $\log_{10}U$ in systems with subsolar metallicity which also exhibit little age dependence; at $Z=0.04$, the $\log_{10}U$ dependence turns over in the oldest systems.

The right-hand panel in \autoref{fig:NII_Halpha_ratio} shows results from G395M/F290LP medium resolution ($R\sim1000$) grating spectra from NIRSpec. We collate $126$ spectra from DJA v4.2 at $5.6<z<6.5$, calculating \fion{N}{2}/\ha\ from Gaussian fits to the emission lines using {\tt specFitMSA}. We find $N_{\rm spec}=\{23,10,3\}$ spectra with \fion{N}{2}$_{\lambda6548}$ and \fion{N}{2}$_{\lambda6584}$ detected at $\{1,2,3\}\sigma$. While we do not compute the ionization parameter for these sources, we find \fion{N}{2}/\ha$~=12.7^{+2.8}_{-0.4}\%$ for the $\mathrm{SNR_{[NII]}}>3$ sample. Relaxing to $\mathrm{SNR_{[NII]}}>\{2, 1\}$, we measure $\{9.1^{+5.6}_{-4.0}, 7.0^{+4.7}_{-6.4}\}\%$. 

Our best-fitting \bagpipes\ templates output this \fion{N}{2}/\ha value, however not enough information is contained within the photometry to constrain this. This makes the inferred values extremely sensitive to systematics in the SED fitting process which are already substantial for measuring the much stronger $\rm H\alpha$ line fluxes/EWs. Since what is important is how much \bagpipes\ gets this value wrong by, and these measured ratios are on average $\simeq10\%$, we ignore the impact of this on our measured \xiionzero. 

\FloatBarrier

\bibliography{main}{}
\bibliographystyle{aasjournalv7}



\end{document}